\documentclass[fleqn,usenatbib]{mnras}

\usepackage{newtxtext,newtxmath}
\usepackage[T1]{fontenc}

\DeclareRobustCommand{\VAN}[3]{#2}
\let\VANthebibliography\thebibliography
\def\thebibliography{\DeclareRobustCommand{\VAN}[3]{##3}\VANthebibliography}

\usepackage{graphicx}	% Including figure files
\usepackage{amsmath}	% Advanced maths commands
\usepackage{xcolor}
\usepackage{float}
\usepackage{upgreek}
\usepackage{subfig}
\usepackage{bm}
\usepackage{pifont}
\usepackage{orcidlink}
\usepackage{multirow}
\newcommand{\tick}{\ding{51}}
\newcommand{\notick}{\ding{55}}
\usepackage[normalem]{ulem}
\usepackage{cancel}
\usepackage{array}
\usepackage{dcolumn}
\usepackage{caption}
\usepackage{footnote}
\usepackage{natbib}

\title[A Gaussian-processes approach to fitting for time-variable spherical solar wind]{A Gaussian-processes approach to fitting for time-variable spherical solar wind in pulsar timing data}

\author[I. C. Ni\c{t}u et al.]{\parbox{\textwidth}{Iuliana C. Ni\c{t}u\orcidlink{0000-0003-3611-3464},$^{\!1}$\!\thanks{E-mail: iuliana-camelia.nitu@manchester.ac.uk}
\,Michael~J.~Keith\orcidlink{0000-0001-5567-5492},$^{\!1}$\!\thanks{E-mail: michael.keith@manchester.ac.uk}
\,Caterina~Tiburzi,$^{\!2}$
\mbox{\,Marcus~Br\"{u}ggen,$^{\!3}$}
\mbox{\,David~J.~Champion\orcidlink{0000-0003-1361-7723},$^{\!4}$}\\
\mbox{Siyuan~Chen\orcidlink{0000-0002-3118-5963},$^{\!5}$}
\mbox{\,Isma\"{e}l~Cognard\orcidlink{0000-0002-1775-9692},$^{\!6,7}$}
\mbox{\,Gregory~Desvignes\orcidlink{0000-0003-3922-4055},$^{\!4}$}
\mbox{\,Ralf-J\"{u}rgen~Dettmar,$^{\!8}$}\\
\mbox{Jean-Mathias~Grie{\ss}meier\orcidlink{0000-0003-3362-7996},$^{\!6,7}$}
\mbox{\,Lucas~Guillemot\orcidlink{0000-0002-9049-8716},$^{\!6,7}$}
\mbox{\,Yanjun~Guo,$^{\!4}$}
\mbox{\,Matthias Hoeft,$^{\!9}$}
\mbox{\,Huanchen~Hu\orcidlink{0000-0002-3407-8071},$^{\!4}$}\\
\mbox{Jiwoong~Jang\orcidlink{0000-0003-4454-0204},$^{\!4}$}
\mbox{\,Gemma~H.~Janssen\orcidlink{0000-0003-3068-3677},$^{\!10,11}$}
\mbox{\,Jedrzej~Jawor\orcidlink{0000-0003-3391-0011},$^{\!4}$}
\mbox{\,Ramesh~Karuppusamy\orcidlink{0000-0002-5307-2919},$^{\!4}$\,\,\,}\\
\mbox{Evan~F.~Keane\orcidlink{0000-0002-4553-655X},$^{\!12}$}
\mbox{\,Michael~Kramer,$^{\!4}$}
\mbox{\,J\"{o}rn~K\"{u}nsem\"{o}ller,$^{\!13}$}
\mbox{\,Kristen~Lackeos\orcidlink{0000-0002-6554-3722},$^{\!4}$}
\mbox{\,Kuo~Liu\orcidlink{0000-0002-2953-7376},$^{\!14,4}$}\\
\mbox{Robert~A. Main,$^{\!4}$}
\mbox{\,James~W.~{McKee}\orcidlink{0000-0002-2885-8485},$^{\!15,16}$}
\mbox{\,Nataliya~K.~Porayko,$^{\!4}$}
\mbox{\,Golam~M.~Shaifullah\orcidlink{0000-0002-8452-4834},$^{\!2,17,18}$}\\
\mbox{Gilles~Theureau\orcidlink{0000-0002-3649-276X},$^{\!6,7,19}$}
and \mbox{\,Christian~Vocks$^{20}$}
}
\\ \\ \\
\parbox{\textwidth}{$^{1}$Jodrell Bank Centre for Astrophysics, Department of Physics and Astronomy, The University of Manchester, Manchester M13 9PL, UK\\
$^{2}$INAF --- Osservatorio Astronomico di Cagliari, via della Scienza 5, 09047 Selargius (CA), Italy\\
$^{3}$University of Hamburg, Gojenbergsweg 112, 21029 Hamburg, Germany\\
$^{4}$Max-Planck-Institut f{\"u}r Radioastronomie, Auf dem H{\"u}gel 69, 53121 Bonn, Germany\\
$^{5}$Kavli Institute for Astronomy and Astrophysics, Peking University, Beijing 100871, China\\
$^{6}$Laboratoire de Physique et Chimie de l'Environnement et de l'Espace LPC2E UMR7328, Université d'Orléans, CNRS, F-45071 Orléans, France\\
$^{7}$Observatoire Radioastronomique de Nançay,
Observatoire de Paris, Université PSL,
CNRS, Université d’Orléans,
18330 Nançay, France\\
$^{8}$Ruhr University Bochum, Faculty of Physics and Astronomy, Astronomical Institute (AIRUB), 44780 Bochum, Germany\\
$^{9}$Th\"{u}ringer Landessternwarte, Sternwarte 5, 07778, Tautenburg, Germany\\
$^{10}$ASTRON, Netherlands Institute for Radio Astronomy, Oude Hoogeveensedijk 4, 7991 PD, Dwingeloo, The Netherlands\\
$^{11}$Department of Astrophysics/IMAPP, Radboud University Nijmegen, P.O. Box 9010, 6500 GL Nijmegen, The Netherlands\\
$^{12}$School of Physics, Trinity College Dublin, College Green, Dublin 2, D02 PN40, Ireland\\
$^{13}$Fakultät für Physik, Universität Bielefeld, Universitätsstr. 25, 33615, Bielefeld, Germany\\
$^{14}$Shanghai Astronomical Observatory, Chinese Academy of Sciences, 80 Nandan Road, Shanghai 200030, China\\
$^{15}$E. A. Milne Centre for Astrophysics, University of Hull, Cottingham Road, Kingston-upon-Hull, HU6 7RX, UK\\
$^{16}$Centre of Excellence for Data Science, AI and Modelling (DAIM), University of Hull, Cottingham Road, Kingston-upon-Hull, HU6 7RX, UK\\
$^{17}$Dipartimento di Fisica ``G. Occhialini", Universit{\'a} degli Studi di Milano-Bicocca, Piazza della Scienza 3, I-20126 Milano, Italy\\
$^{18}$INFN, Sezione di Milano-Bicocca, Piazza della Scienza 3, I-20126 Milano, Italy\\
$^{19}$Laboratoire Univers et Théories, Observatoire de Paris, Université PSL, Université de Paris Cité, CNRS, F-92190 Meudon, France\\
$^{20}$Leibniz-Institut f\"{u}r Astrophysik Potsdam, 14482 Potsdam, Germany\\
}
}
\date{Accepted XXX. Received YYY; in original form ZZZ}

\pubyear{2024}

\begin{document}
\label{firstpage}
\pagerange{\pageref{firstpage}--\pageref{lastpage}}
\maketitle

\begin{abstract}
Propagation effects are one of the main sources of noise in high-precision pulsar timing. For pulsars below an ecliptic latitude of~$5^\circ$, the ionised plasma in the solar wind can introduce dispersive delays of order $100\,\upmu\mathrm{s}$ around solar conjunction at an observing frequency of 300\,MHz.
A common approach to mitigate this assumes a spherical solar wind with a time-constant amplitude. However, this has been shown to be insufficient to describe the solar wind. We present a linear, Gaussian-process piecewise Bayesian approach to fit a spherical solar wind of time-variable amplitude, which has been implemented in the pulsar software \textsc{run\_enterprise}. Through simulations, we find that the current EPTA+InPTA data combination is not sensitive to such variations; however, solar wind variations will become important in the near future with the addition of new InPTA data and data collected with the low-frequency LOFAR telescope. We also compare our results for different high-precision timing datasets (EPTA+InPTA, PPTA, and LOFAR) of three millisecond pulsars (J0030$+$0451, J1022$+$1001, J2145$-$0450), and find that the solar-wind amplitudes are generally consistent for any individual pulsar, but they can vary from pulsar to pulsar. Finally, we compare our results with those of an independent method on the same LOFAR data of the three millisecond pulsars. We find that differences between the results of the two methods can be mainly attributed to the modelling of dispersion variations in the interstellar medium, rather than the solar wind modelling.
\end{abstract}


%%%%%%%%%%%%%%%%%%%%%%%%%%%%%%%%%%%%%%%%%%%%%%%%%%

%%%%%%%%%%%%%%%%% BODY OF PAPER %%%%%%%%%%%%%%%%%%
% for terminal font: \texttt{mnras\_sample.tex}

% for in-text: \citet{Fournier1901},
% in brackets: \citep[e.g.][]{vanDijk1902}.

% equation~(\ref{eq:quadratic})
% Fig.~\ref{fig:example_figure}
% Table~\ref{tab:example_table}

%\begin{figure}
%	\includegraphics[width=\columnwidth]{example}
%    \caption{}
%    \label{fig:example_figure}
%\end{figure}

%\begin{table}
%	\centering
%	\caption{}
%	\label{tab:example_table}
%	\begin{tabular}{lccr} % four columns, alignment for each
%		\hline
%		A & B & C & D\\
%		\hline
%		1 & 2 & 3 & 4\\
%		2 & 4 & 6 & 8\\
%		3 & 5 & 7 & 9\\
%		\hline
%	\end{tabular}
%\end{table}
\nopagebreak
\nobreak
\section{Introduction}

Pulsar timing consists in recording the times of arrival (ToAs) of highly stable pulses emitted by pulsars, which are then compared with predictions from long term models. These models take into account the pulsar's behaviour, as well as factors such as astrometric effects (e.g. spin frequency and its derivatives, position, etc.), binary companions, or the dispersive delays induced by the ionised medium through which the pulsar signal propagates \citep[for more details see e.g.][]{Edwards2006}. The left-over signal after subtracting the model from the observed ToAs is often referred to as `timing residuals', or just `residuals', and is expected to be only white noise if the model is optimal.

Due to the high stability of pulsar rotation, and particularly of the population of recycled millisecond pulsars \citep[MSPs;][]{Backer1982}, high-precision pulsar timing constitutes a great tool for a large variety of scientific investigations \citep[see e.g.][]{Manchester2017}. Huge efforts are currently being concentrated on gravitational wave searches using decades of observations of a large sample of MSPs with multiple telescopes, referred to as Pulsar Timing Arrays \citep[PTAs; e.g.][]{Tiburzi2018}. PTA experiments are expected to be primarily sensitive to the gravitational wave background (GWB) in the nanohertz-frequency regime, most likely originating from supermassive black hole binary mergers \citep[e.g.][]{BurkeSpolaor2019}. Three major collaborations have historically been involved in the search for the GWB, namely the European Pulsar Timing Array \citep[EPTA;][]{Desvignes2016}, the Parkes Pulsar Timing Array \citep[PPTA;][]{Manchester2013}, and the North American Nanohertz Observatory for Gravitational Waves \citep[NANOGrav;][]{Demorest2013}; these are also the founding members of the International Pulsar Timing Array \citep[IPTA;][]{Verbiest2016} collaboration, having recently been joined by the Indian Pulsar Timing Array \citep[InPTA;][]{Joshi2022}. Recently, the three aforementioned collaborations (with the InPTA working along the EPTA) all coherently reported marginal evidence for a GWB signal \citep{EPTA2023III,Reardon2023,Agazie2023}. While neither of these currently meet the requirements for being defined as a clear detection, further  investigations into the analysis methods, as well as the upcoming combined IPTA dataset, are expected to improve on these GWB measurements.

The signal induced by the GWB in pulsar data is expected to be extremely weak, even when correlated over tens of pulsars \citep[e.g.][]{Siemens2013,Janssen2015}. Therefore, other effects in pulsar timing data must be carefully considered, as they can obscure, or even mimic, a GWB signal \citep{Tiburzi2016}. One of the strongest sources of `noise' in this context are the dispersive delays introduced in the ToAs by the interaction between the radio waves and the ionised medium through which the pulsar signal propagates on its way to the observer \citep{Lentati2016}. These dispersive delays are modelled as having an inverse-squared dependency with the observing frequency, $f_\mathrm{obs}$. The dispersive delay on a ToA is expressed as
\begin{equation}
    \label{eq: tDM}
   {t^{}_\mathrm{D}} = \dfrac{1}{K_\mathrm{\!D} f_\mathrm{obs}^2} \int\limits_{\ell} \!\!\!n_\mathrm{e} \mathrm{d}\ell = \dfrac{1}{K_\mathrm{\!D} f_\mathrm{obs}^2} \, {\mathrm{{DM}}}, 
\end{equation}
where $K_\mathrm{\!D} \simeq 2.41\times 10^{-4} \,\mathrm{MHz}^{-2}\ \mathrm{pc} \,\mathrm{cm}^{-3}\,\mathrm{s}^{-1}$ is a dispersion constant \citep{Manchester1972}, and the dispersion measure DM is defined as the free electron number density, $n_\mathrm{e}$, integrated over the line of sight to the pulsar, $\ell$; the DM is usually quoted in $\mathrm{pc}\,\mathrm{cm}^{-3}$.
As illustrated by Eq.~\ref{eq: tDM}, the dispersion delay is stronger at lower observing frequencies. The noise due to the dispersion delay is dominated by the effects of the turbulent and inhomogeneous ionised interstellar medium (IISM) along the line-of-sight, which can induce fluctuations in the DM of order $10^{-3}\,\mathrm{pc\,cm}^{-3}$ over a timescale of years \citep[e.g.][]{Keith2013,Jones2017,Donner2020}. Several possible mitigating strategies for the turbulent IISM contribution are used throughout the PTAs, such as modelling it as a chromatic red-noise Gaussian process in Fourier space \citep[e.g.][]{Lentati2014}, or using a time-domain piecewise binned model \citep[e.g. the DMX model;][]{Arzoumanian2015}. 

For pulsars with a line-of-sight that passes close to the Sun, the delay induced by the propagation through the solar wind (SW hereafter) is also noticeable in the current quality of pulsar data, inducing DM fluctuations as high as $10^{-3}\text{--}10^{-4}\,\mathrm{pc\,cm}^{-3}$. Furthermore, \citet{Tiburzi2016} showed that, if not carefully considered, the influence of the SW may mimic a GWB signal in PTA-like data, as the SW can create spatial correlations among the pulsars. To account for the influence of the SW in pulsar timing data, PTA collaborations generally use a simple `spherical SW' model, based on the assumption that the number density of ionised electrons varies under a spherically symmetric law away from the Sun, according to the inverse square law \citep{Edwards2006}, i.e. 
\begin{equation}
    \label{eq: nesph}
    n_\mathrm{e} (\bm{r}) = N_\mathrm{e}^\mathrm{SW} \left(\dfrac{1\,\mathrm{au}}{|\bm{r}|}\right)^2,
\end{equation}
where $\bm{r}$ is the position vector from the Sun to the point of interest affected by the SW, generally given in astronomical units (au); $N_\mathrm{e}^\mathrm{SW}$ is the amplitude of the number density at 1\,au, and we generally refer to it as the `amplitude of the SW' in this work. Under the spherical SW assumption, this amplitude $N_\mathrm{e}^\mathrm{SW}$ is space invariant (does not depend on the vector $\bm{r}$); thus integrating the number density as per Eq.~\ref{eq: nesph} over the line-of-sight gives the DM contribution as \citep[e.g.][]{You2007b}
\begin{equation}
    \label{eq: DMsphsw}
    \mathrm{DM}^\mathrm{SW}_\mathrm{sph} \simeq 4.85 \times 10^{-6} \, \bigg(\dfrac{N_\mathrm{e}^\mathrm{SW}}{\mathrm{cm}^{-3}}\bigg) \bigg(\dfrac{\pi-\theta}{\sin{\theta}}\bigg) \,\,\mathrm{pc}\,\mathrm{cm}^{-3},
\end{equation}
where $\theta$ is the solar elongation angle of the pulsar, i.e. the pulsar-observer-Sun angle, which is minimum at the solar conjunction of the pulsar. Replacing the DM given as per Eq.~\ref{eq: DMsphsw} into Eq.~\ref{eq: tDM} gives the time delay of a ToA, due to a spherical SW, which we factorise as
\begin{equation}
    t_\mathrm{D}^\mathrm{SW} = \dfrac{1}{K_\mathrm{\!D} f_\mathrm{obs}^2} \, N_\mathrm{e}^\mathrm{SW}  S_{\!\theta},
\end{equation}
where we have summarised some of the physical constants and geometrical dependence in the variable
\begin{equation}
    \label{eq: Ssph}
    S_{\!\theta} \simeq  4.85 \times 10^{-6} \,\bigg(\dfrac{\pi-\theta}{\sin{\theta}}\bigg)\,\,\mathrm{pc} \simeq  \bigg(\dfrac{\pi-\theta}{\sin{\theta}}\bigg)\,\,\mathrm{au}.
\end{equation}

In the standard approach to mitigating the effect of the SW, the amplitude $N_\mathrm{e}^\mathrm{SW}$ in the described spherically symmetric model is constant in time; in e.g. the recent EPTA dataset \citep{EPTA2023I}, this amplitude is generally kept fixed at $7.9\,\mathrm{cm}^{-3}$, as per \citet{Madison2019}. Furthermore, data taken when the pulsar appears $< 5^{\circ}$ away from the Sun in the sky are considered to be poorly described by this simple model and commonly removed \citep{Verbiest2016}. 

No model has yet been developed that fully captures the observed impact of the SW, especially on low-frequency pulsar data, as shown by e.g. \citet{Tiburzi2019}. \citet{You2007b} proposed a model based on the coronal magnetograms derived by the Wilcox Solar Observatory, and the bi-modal nature of the SW, considering the contributions from both a slow (equatorial) and a fast (polar) solar stream \citep[e.g.][]{Coles1996}. \citet{You2007b} argued that this two-phase model performed better than the spherical SW for their PPTA observations. However, \citet{Tiburzi2019} found that the spherically symmetric SW model with time-dependent amplitude performed systematically better than the two-phase model in removing the SW contribution in their longer-span and lower-frequency LOFAR data. The apparent discrepancy between these two analyses is thought to be due to either (i) the increased DM precision of the lower observing frequency of the data used in \citet{Tiburzi2019} as compared with that of \citet{You2007b}; or (ii) a difference in performance of the two-phase model with the heliospheric latitude of the pulsar, since the two papers investigated data from different pulsars.

A clear improvement to current general models is to allow the amplitude of the SW to vary each year (i.e. be time-variable) in the spherically symmetric model. \citet{Tiburzi2021}, hereafter referred to as T21, have already shown this to be beneficial and more adequate for low-frequency data taken with the European interferometer LOFAR \citep{vHaarlem2013}. T21 fit a time-variable spherical SW to each pulsar, and a clear temporal variation was observed in several of the analysed pulsars. Note that previously, \citet{Madison2019} had reported little evidence for long-term variations in the SW density using the NANOGrav 11-yr dataset.\\

In this work, we present a Gaussian-process piecewise approach, implemented as part of the pulsar analysis toolkit \textsc{run\_enterprise} \citep{run_enterprise}, which allows for an automatic time-variable spherical SW fitting in a Bayesian framework, simultaneously with all the other pulsar timing parameters and noise models.
Recently, \citet{Hazboun2022} developed several comprehensive Bayesian algorithms that allow fitting for SW across several pulsars simultaneously. They used a uniform-prior piecewise model that globally fits for a spherical SW amplitude in each temporal bin (of e.g. 3 months) of the data, also exploring variations in the exponent of the $1/r^2$ law of Eq.~\ref{eq: nesph}; further, they also explore a model based on globally fitting for continuous Fourier-basis, time-dependent variations in the SW. In this work, we present a different approach to SW Bayesian fitting. We use a simple piecewise algorithm based on a Gaussian process with a Normal-distribution prior to fit for a time-variable amplitude spherical SW in each pulsar individually. The choice of prior allows us to marginalise over the individual yearly amplitudes, reducing the problem to a single additional hyperparameter in the width of the Gaussian prior. This, together with the mathematical implementation that keeps the parameter estimation algebraically linear, makes any additional computational cost negligible.

This article is structured as follows: in Section~\ref{sec: observations}, the main properties of the datasets are summarised; in Section~\ref{sec: method} we describe the algorithm implemented in our pipeline and simulations; in Section~\ref{sec: resultsdiscuss} the results of our analysis are discussed; and in Section~\ref{sec: conclusions} we summarise our conclusions.

\iffalse
\citet{Tiburzi2016, Tiburzi2019, Tiburzi2021} , 
\citet{You2007a, You2007b, You2012}
\citet{Madison2019}  
PPTA (and FAST) SW paper: \citet{Niu2017}  
GMRT SW paper: \citet{Krishnakumar2021}  
NANOGrav Bayesian SW paper: \citet{Hazboun2022}
\citet{Kumar2022}
\fi

\section{Datasets} \label{sec: observations}
For this work, we selected three millisecond pulsars which pass in close proximity ($\leq\!\!5.31^{\circ}$) to the Sun during conjunction, and are included in the study by T21. These are PSRs J0030$+$0451, J1022$+$1001, and J2145$-$0450, and they are present both in LOFAR and PTA observations. Here we consider data from the recent EPTA+InPTA data combination \citep{EPTA2023I, Tarafdar2022}, as well as from the Second Data Release (DR2) of PPTA \citep{Reardon2021}. Table~\ref{tab: realdataused} shows the ecliptic latitudes of the three pulsars of interest, equivalent to the sky-angle between the pulsar and the Sun at conjunction. It also shows which data were available for each pulsar.
%\protect\footnote{The ecliptic latitudes are quoted from the ATNF Pulsar Catalogue, https://www.atnf.csiro.au/research/pulsar/psrcat/}
\begin{table}
	\centering
	\caption{Ecliptic latitudes (elat) and summary of which datasets are available for the three pulsars used in this work; a tick (\tick) signifies the data are available, whereas a cross (\notick) means they are not. By EPTA availability we refer specifically to the Data Release 2 dataset. The EPTA and InPTA data are combined for PSR J1022$+$1001.}
	\label{tab: realdataused}
	\begin{tabular}{c||D{.}{.}{2}||rc|c|c} % four columns, alignment for each
		\hline
		  PSR & \multicolumn{1}{c||}{elat [$^{\circ}$]} & \multicolumn{2}{c|}{EPTA + InPTA} & PPTA & LOFAR\\
		\hline
		\hline
		J0030$+$0451 & \,\,\,\,\,$1.45$ & \,\,\,\,\,\tick & \notick & \notick & \tick \\
		  J1022$+$1001 & $-0.06$ & \,\,\,\,\tick & \tick & \tick & \tick \\
		J2145$-$0450 & $5.31$ & \,\notick & \tick & \tick & \tick \\
		\hline
	\end{tabular}
\end{table}
The datasets, each with different frequency coverage and properties (discussed in more detail in Sections~\ref{sec: EPTA+InPTA}-\ref{sec: PPTA}), are used separately to construct simulations, and thus assess their sensitivity to the SW. Further, the results of our pipeline are compared across all the datasets, as well as with the independent measurements in T21.

\subsection{EPTA (DR2) + InPTA (DR1) combined dataset} \label{sec: EPTA+InPTA}

The Second Data Release (DR2) of the EPTA collaboration is used in this work, augmented with the First Data Release (DR1) of the InPTA \citep{EPTA2023I, Tarafdar2022}. Here, we refer to this combined dataset as EPTA+InPTA.

There are five European radio telescopes which provided data for the EPTA DR2 dataset, namely: the 100-m Effelsberg Telescope (in Germany), Jodrell Bank Observatory's 76-m Lovell Telescope (in the United Kingdom), Nan\c{c}ay Radio Observatory's large Radio Telescope (NRT; in France), the Astronomical Observatory of Cagliari's 64-m Sardinia Radio Telescope (SRT; in Italy), and the Westerbork Synthesis Radio Telescope (WSRT; in the Netherlands). Moreover, these telescopes were also used collectively, on a monthly cadence, as the Large European Array for Pulsars (LEAP), equivalent to a 194-m sixth interferometric telescope in the EPTA \citep{Bassa2016}. The EPTA DR2 dataset contains observations of 25 millisecond pulsars, up to 25 years in length. The large majority of these observations were taken at frequencies in the `L-band' ($1\text{--}2\,\mathrm{GHz}$) and above, with bandwidths of up to $512\,\mathrm{MHz}$, while there were a limited number of observations centred at lower frequencies of $350\,\mathrm{MHz}$. For a detailed description of the properties of the EPTA DR2 telescopes and observations, see \citet{Chen2021}, and \citet{EPTA2023I}.

The InPTA dataset includes observations taken with the upgraded Giant Meterwave Radio Telescope (uGMRT; in India) over a period of 3.5 years \citep[for a detailed description see][]{Tarafdar2022}. The uGMRT took simultaneous observations at two frequency bands, referred to as `B3' ($300\text{--}500\,\mathrm{MHz}$), and `B5' ($1260\text{--}1460\,\mathrm{MHz}$), respectively. These simultaneous observations at different frequencies, as well as the wide bandwidths, make the InPTA observations extremely valuable for measuring the DM influence in pulsar data, including the effect of the SW.

\subsection{LOFAR dataset} \label{sec: LOFAR}

For PSRs J0030$+$0451 and J2145$-$0750, the same LOFAR datasets as in T21 are used; while for PSR J1022$+$1001, the T21 dataset was supplemented with the most recent $\sim2$ years of observations for this work. These data were taken with subsets of the International LOFAR telescope \citep{vHaarlem2013,Stappers2011}, namely six of the German stations, the Swedish station, and the LOFAR Core in the Netherlands. More than 100 pulsars are regularly observed using this setup. The extremely low observing frequency of the LOFAR instruments, covering a range of roughly $110\text{--}190\,\mathrm{MHz}$, makes this telescope incredibly well-suited for studying chromatic effects on pulsar data such as the SW. For more details on the LOFAR datasets, see e.g. \citet{Porayko2019}, \citet{Donner2019}, \citet{Tiburzi2019}.

\subsection{PPTA (DR2) dataset} \label{sec: PPTA}

We also make use of the open-access PPTA DR2 timing data \citep{Reardon2021} to validate and compare with the EPTA+InPTA results. This PPTA dataset is described in detail in \citet{Kerr2020} and was taken with the Australian 64-m Murriyang Parkes Radio Telescope. It spans $\sim\!\!14$ years ($2004\text{--}2018$) of observations of 26 millisecond pulsars at three frequency bands, roughly centred at $700, 1400$ and $3100\,\mathrm{MHz}$. The observing cadence for each pulsar was approximately two-three weeks. The ToAs are available in both sub-banded and band-averaged form, of which we used the latter in this work, for simplicity. In particular, the PPTA DR2 data for PSRs J1022$+$1001 and J2145$-$0450 were used to compare SW results with those of the LOFAR and EPTA+InPTA ToAs. The data of PSR J2145$-$0450 was also supplemented with archival observations of the Parkes telescope since 1994, as published in \citet{Manchester2013}; these were however much less sensitive to the SW.

\section{Method} \label{sec: method}
\subsection{Modelling} \label{sec: modelling}
\subsubsection{Timing model}

In this work, we use open-source pulsar software, namely the Bayesian fitting software \textsc{enterprise} \citep{Ellis2019}, which has been integrated into the pulsar analysis toolkit \textsc{run\_enterprise} \citep{run_enterprise}, together with the pulsar timing software \textsc{tempo2} \citep{Edwards2006}.

The Bayesian method allows for simultaneous fitting of deterministic pulsar parameters (such as the spin frequency and its derivatives, position, known binary companions, etc.), which are generally marginalised over, unless of particular interest; and of parameters characterising stochastic timing noise. In this framework, white noise in timing data is described by fitting the parameters `EFAC' and `EQUAD' which can rescale the ToA error bars, and account for additional white noise, respectively \citep{Edwards2006}. Red timing noise is described using Fourier-domain Gaussian-process modelling as described in \citet{Lentati2014}. The sinusoidal Fourier-basis components at each Fourier frequency $f$ are multiplied by a set of amplitudes described by a Gaussian process; we refer to the covariance of these amplitudes as the power spectral density. In our fitting, we model the power spectral density prior as a power-law, as often used in the pulsar community \citep{Lentati2014,vanHaasteren2011}. However, for this study we choose to characterise this power-law by an amplitude at a reference frequency of $0.1\,\mathrm{yr}^{-1}$, different from the usual choice of $1\,\mathrm{yr}^{-1}$; this choice ensures the amplitude is more robust against the yearly periodicity of the solar wind. We therefore characterise the power spectral density model as
\begin{equation}
    \label{eq: psd}
    P(f) = \mathcal{A} \left(\dfrac{f}{0.1\,\mathrm{yr}^{-1}}\right)^{-\gamma} \mathrm{yr}^3,
\end{equation}
where $\mathcal{A}$ and $\gamma$ correspond to the Bayesian hyperparameters characterising the red timing noise, while the individual Gaussian-process amplitudes are marginalised over. This is used for both achromatic red timing noise, and chromatic DM noise, which has an inverse-square dependence on observing frequency, as per Eq.~\ref{eq: tDM}. For the former, the power amplitude is often written as $\mathcal{A} \!\equiv\! A_\mathrm{red}^2/(12\pi^2)$, and $\log_{10} (A_\mathrm{red})$ is chosen as the fitting hyperparameter together with the exponent $\gamma\!\equiv\!\gamma_{\mathrm{red}}$. For the $\mathrm{DM}$ noise, the Fourier components are also proportional to the square-inverse of the observing frequency $f_\mathrm{obs}$; the prior power amplitude is expressed as $\mathcal{A} \!\equiv \!A_\mathrm{DM}^2$, and $\log_{10}(A_\mathrm{DM})$ and $\gamma\!\equiv\!\gamma_{\mathrm{DM}}^{}$ are the fitting hyperparameters.

\subsubsection{Solar wind fitting using Gaussian processes} \label{sec: swfitting}

We present an implementation of time-variable amplitude spherical SW modelling which has been incorporated in the \textsc{run\_enterprise} package and can therefore be included in the simultaneous Bayesian fitting of pulsar timing data, together with the other deterministic pulsar parameters, and the white and red noise parameters.

\begin{figure}
	\includegraphics[width=\columnwidth]{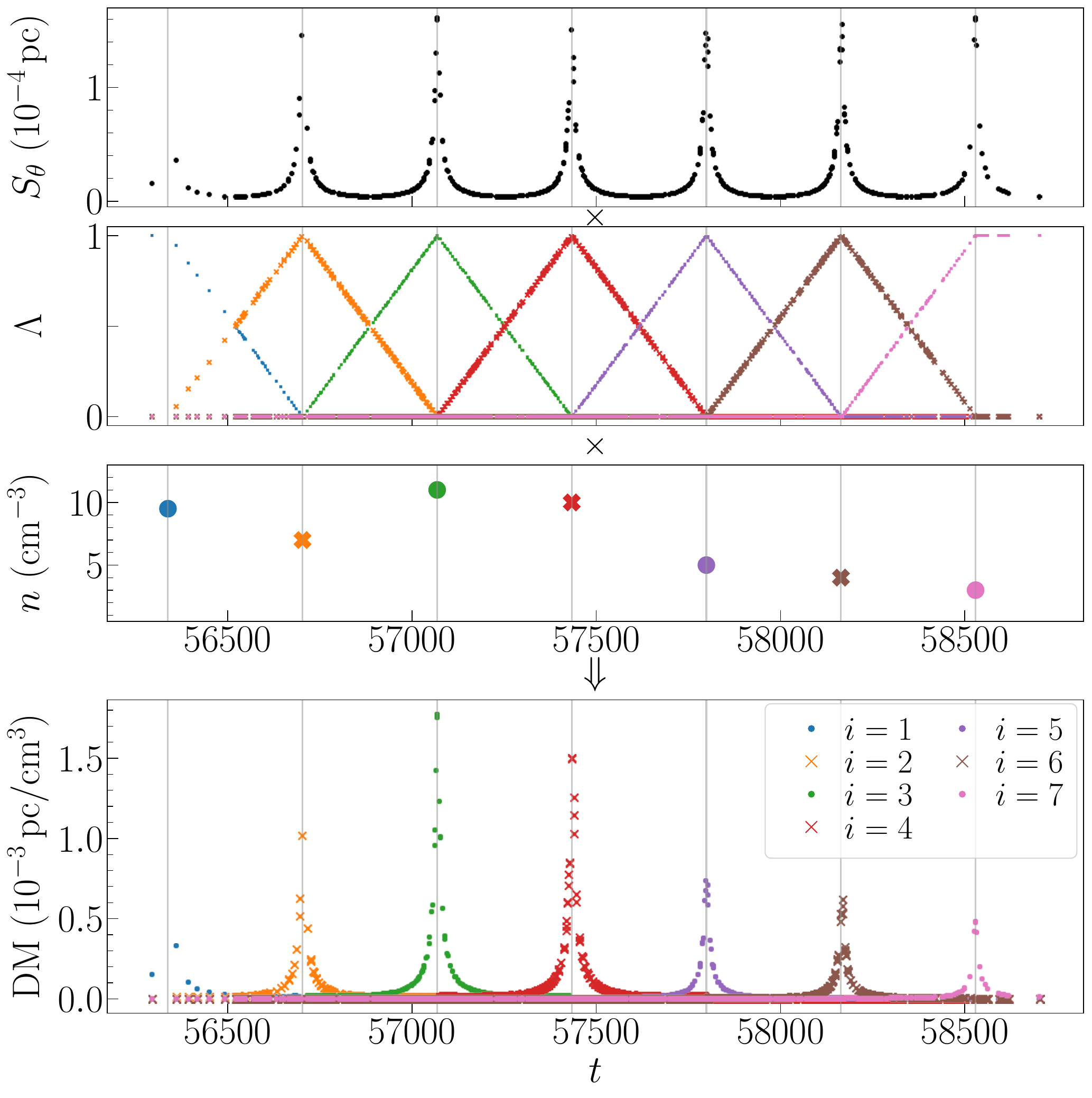}
    \caption{Graphical illustration of the mathematical representation in Eqs.~\ref{eq: LN}--\ref{eq: lambda}. The quantities in the top three plots ($S_{\theta}$, $\Lambda$, and $n$) are multiplied to give the DM (bottom plot). Note that we do not show the dependence on observing frequency for simplicity, and $\mathrm{DM} = K_\mathrm{\!D} f_\mathrm{obs}^2 \,\bm{n} \mathbf{V}$. Each colour represents a different component, corresponding to each solar conjunction, $T_{\!i}$. The data points represent the ToAs $t_{\!j}$.}
    \label{fig: SWfitting}
\end{figure}

An approach based on Gaussian processes is used, similar in concept to that used to describe the red noise model as presented in \citet{Lentati2014}. The SW signal is expressed, in the time domain, as the sum of independent components, which correspond to the unit-amplitude spherical SW for each solar conjunction, multiplied by a set of amplitudes equivalent to the quantity $N_\mathrm{e}^\mathrm{SW}$ from Eq.~\ref{eq: DMsphsw}. We choose a set of simple piecewise linear (`triangular') functions, with centre points at the solar conjunction times of the pulsar, to mathematically represent each yearly SW variation; this follows the idea presented in \citet{Keith2013} for modelling DM noise in general. In practice, the SW contribution to the vector of ToAs is written as
\begin{equation}
    \label{eq: LN}
    \bm{t}_\mathrm{sw} = \bm{n} \mathbf{V},
\end{equation}
where $\bm{n}$ is a column vector containing the spherical SW amplitudes equivalent to $N_\mathrm{e}^\mathrm{SW}$ for each of the $N_\mathrm{c}$ solar conjunctions within the data. $\mathbf{V}$ is a matrix of size $(N_\mathrm{c} \times N_\mathrm{t})$, where $N_\mathrm{t}$ is the number of ToAs, i.e. the size of the column vector $\bm{t}_\mathrm{sw}$. We define an element of the matrix $\mathbf{V}$,  such that
\begin{equation}
    \mathrm{V}_{\!i\!j} = \dfrac{1}{K_\mathrm{\!D}f_\mathrm{obs}^2} \, {S}_{\!\theta\!, j} \, \Lambda\bigg(\!\dfrac{t_{\!j} - T_{\!i}}{1\,\mathrm{yr}}\!\bigg), 
\end{equation}
where $i\!\in\! \{1, \dots, N_\mathrm{c}\}$ is the matrix row and $j\!\in\! \{1, \dots, N_\mathrm{t}\}$ the column; ${S}_{\!\theta\!, j}$ is defined as per Eq.~\ref{eq: Ssph} for the solar elongation angle $\theta_{\!j}$ corresponding to the ToA $t_{\!j}$. $T_{\!i}$ is the time of the $i$th solar conjunction in the data, and $\Lambda$ is the triangular function, such that by definition
\begin{equation}
\label{eq: lambda}
    \Lambda(q) = \left\{
\begin{array}{ll}
      1 - |q|,& |q| < 1; \\
      0,& \mathrm{otherwise}.
\end{array} 
\right.
\end{equation}
Fig.~\ref{fig: SWfitting} illustrates the steps in creating the SW components in this model.

We assume the distribution of SW amplitudes to be characterised by a Gaussian function, of mean value $N_0$, and a standard deviation $\sigma_\mathrm{sw}$. 
In practice, we therefore express the vector of amplitudes $\bm{n}$ as a sum
\begin{equation}
    \bm{n} = N_0 \mathrm{\textit{\textbf{1}}}  + \bm{\delta\!n},
\end{equation}
where $N_0$ is the mean SW amplitude (i.e. the equivalent of the standard \textsc{tempo2} parameter `NE\_SW'), and $\mathrm{\textit{\textbf{1}}}$ is an $N_\mathrm{c}$-point column vector of ones. $\bm{\delta\!n}$ is therefore the column vector of SW amplitude variations away from the mean. We assume the elements of $\bm{\delta\!n}$ to have mean zero, and their variation with each solar conjunction to be characterised by a Gaussian process, and therefore by a corresponding covariance matrix. In this analysis, we use a simple constant-variance, such that the covariance matrix of the SW amplitude variations $\bm{\delta\!n}$ is a diagonal matrix with all diagonal elements equal to the variance $\sigma_\mathrm{\!sw}^2$.
The quantity $\sigma_\mathrm{\!sw}^{}$ therefore sets the prior for the amplitudes $\bm{\delta\!n}$, and is a hyperparameter referred to as `SW\!\_sigma' in the fitting code, and in the output parameter file. Note that one can also choose a different, more complicated covariance informed by physical processes, e.g. one that follows the 11-yr solar cycle. While this may be explored in the future, we believe that the simple approach taken here is suited for capturing the year-by-year variability in amplitude in the current datasets. 

Similarly to the approach taken in Fourier-basis red-noise fitting, we marginalise over the Gaussian-process amplitudes $\bm{n}$ and only keep $\sigma_\mathrm{\!sw}^{}$ as a Bayesian fitting hyperparameter. As this model keeps the parameter estimation algebraically linear, including the additional time-variable spherical SW model when fitting a pulsar dataset is computationally inexpensive, as it only adds one extra hyperparameter to the entire timing model. Furthermore, \textsc{tempo2} can be used after the Bayesian fit performed with the \textsc{run\_enterprise} software package to explicitly find the amplitudes of the spherical SW at each solar conjunction, if of interest. This is done using the implemented constraining of the least-squares fitting, which takes into account the fitted $\sigma_\mathrm{\!sw}^{}$; for an explanation of how this works, see Appendix~A in \citet{Keith2013}.

\subsection{Simulations}

\subsubsection{General setup}
To test the capabilities of the SW model presented in this work, we created sets of simulated ToAs of both uniform and PTA-like cadence. For all simulations, we started from a set of `idealised' ToAs (characterised by zero residuals) for a chosen pulsar, and added realistic levels of noise, informed by typical values in the observations available. This was done using \textsc{tempo2} plugins, which generate the types of signal discussed in Section~\ref{sec: modelling}, namely:
\begin{itemize}
    \item \textsc{addGaussian} to add white noise;
    \item \textsc{addRedNoise} to add achromatic red noise, characterised by a power-law prior of user-specified amplitude and slope;
    \item \textsc{addDmVar} to add DM noise, also characterised by a power-law prior of user-specified amplitude and slope;
    \item \textsc{addArbitraryDM} to add the DM influence from a simulated SW (see Section~\ref{sec: simSW} for more details on simulating realistic SW DM series).
\end{itemize}
Using the above types of signals, multiple ToA sets were produced, serving various testing purposes which we describe in Section~\ref{sec: resultsdiscuss}. Multiple Gaussian-process realisations were created for the same characteristic hyperparameters, such that we were also able to check the robustness of our pipeline with repeat measurements.

\subsubsection{Simulating the solar wind} \label{sec: simSW}

We simulate the SW influence on a pulsar as a DM time-series that is then added to the total simulated ToAs using the \textsc{tempo2} plugin \textsc{addArbitraryDM}. We consider the variation with respect to the unit-amplitude ($N_\mathrm{e}^\mathrm{SW}\!=\!1$) spherical SW, i.e. the ratio of the DM series to ${S}_{\!\theta}$, and sample it as a Gaussian process characterised by an exponential-squared kernel \citep[see e.g.][]{GP}, of the form
\begin{equation}
    \label{eq: expsqkernel}
    k(\tau) = A_{\!k} \, \exp\left({-\dfrac{\tau^2}{2\lambda^2}}\right),
\end{equation}
where $\tau$ represents the `distance' between two observing times, $A_{\!k}$ is the kernel amplitude, and $\lambda$ is the metric, i.e. the scale of the correlation within the signal described by this kernel. The exponential-squared kernel is a somewhat arbitrary choice, but this simple stationary kernel is widely used and well-suited for describing smooth functions characterised by a single overall metric. In practice, we choose $\lambda=400\,\mathrm{d}$; the exact value of this is not strictly relevant, but a value slightly above $1\,\mathrm{yr}$ ensures that the simulated signal at each solar conjunction is roughly independent of that at the other solar conjunctions. 

The amplitude $A_{\!k}$ broadly characterises the overall variation away from the mean of the signal. To find a realistic value of $A_{\!k}$, we use the SW DM series of PSR J0034$-$0534 as presented in Fig. 3 of T21. We refer the reader to \citet{Tiburzi2019,Tiburzi2021} for details on how this was obtained. In short, using $5\,\mathrm{yr}$ of LOFAR observations, the total DM contribution at each average epoch was estimated. The contribution of the IISM to this DM series was then modelled by a cubic spline with each piece corresponding to a solar conjunction; the SW contribution was simultaneously modelled assuming a spherical SW as in Eq.~\ref{eq: DMsphsw}, while the amplitude was allowed to vary year-by-year. In this work, the DM series after subtracting the cubic spline of the IISM model was used, and we refer to this as ${\mathrm{DM}}^\mathrm{SW}_\mathrm{Lo}$. We define a time series ${y}_0 \equiv {\mathrm{DM}}^\mathrm{SW}_\mathrm{Lo} \!/ {S}_{\!\theta}$, such that ${y}_0$ only encompasses the estimated variation of the SW with respect to a spherical SW of time-invariant, unit-amplitude. Note that at a time $T_{\!i}$ of the $i$th solar conjunction in the data, the value of ${y}_0$ is equivalent to the SW amplitude $n_{i}$. The time series ${y}_0$ is then assumed to be described by a smooth function, characterised by an exponential-squared kernel as in Eq.~\ref{eq: expsqkernel}, of metric $\lambda = 400\,\mathrm{d}$. To find the representative amplitude $A_{\!k}$, we employ the \textsc{python} library \textsc{george} to fit this exponential-squared kernel to the ${y}_0$ series, and find a maximum-likelihood value of $A_{\!k} \approx 3.7\,\mathrm{cm}^{-3}$. 

\begin{figure}
	\includegraphics[width=\columnwidth]{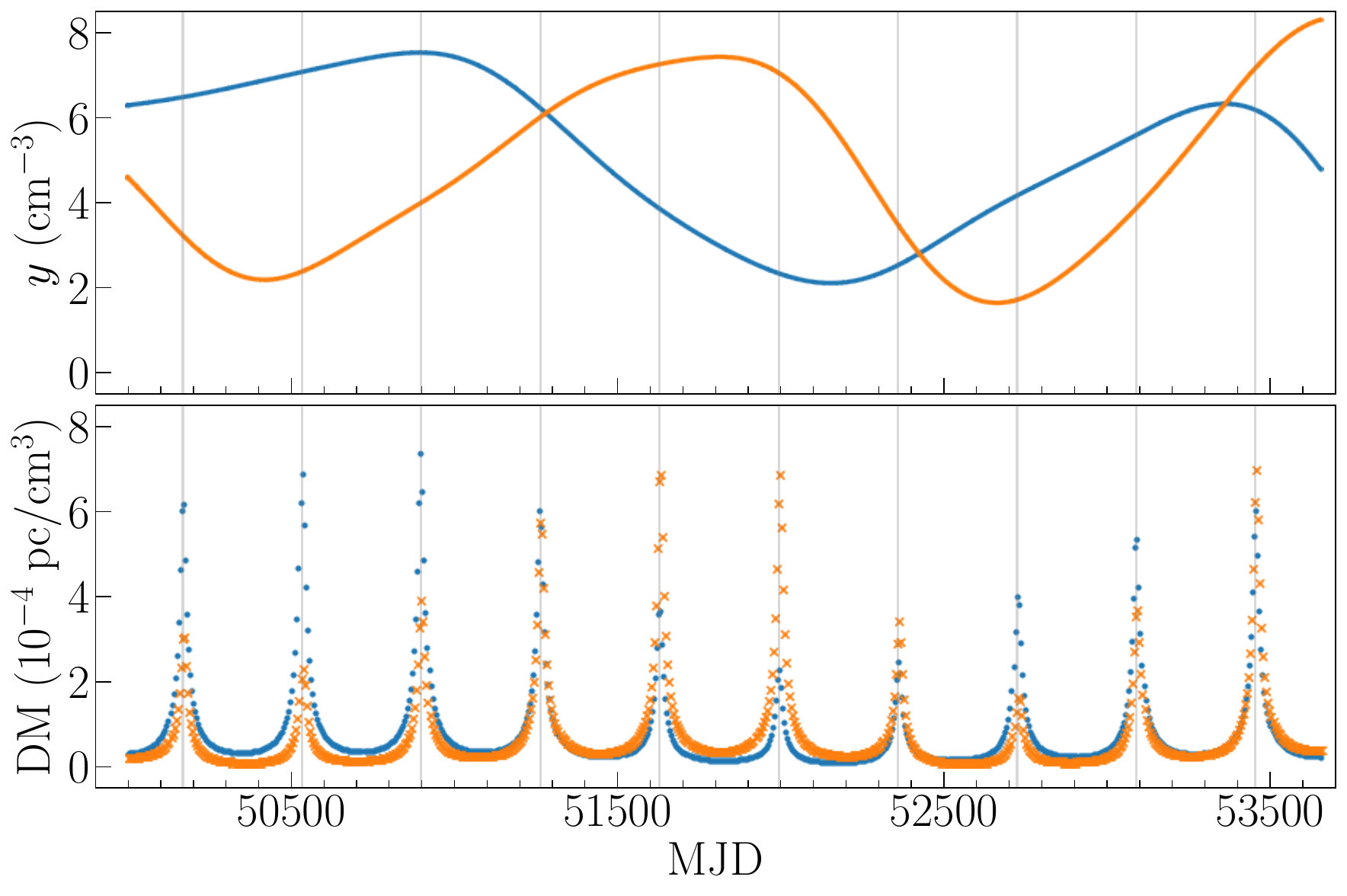}
    \caption{An example of two realisations of simulated SW using an exponential-squared kernel with amplitude informed by real data. The data points are at uniformly sampled ToA values. The top plot represents the simulated series ${y}$, which is then multiplied by the pulsar-specific geometrical factor $S_{\!\theta}$ to create a DM series, shown in the bottom plot.}
    \label{fig: example_DMSW}
\end{figure}

With the fully parameterised exponential-squared kernel, we are able to draw as many new samples of time-series ${y}$ (of the same type as ${y}_0$) as needed, for any pulsar with an available ephemeris. The unit-amplitude spherical SW (i.e. ${S}_{\!\theta}$) of any known pulsar is easily obtainable from the \textsc{tempo2} plugin `\textsc{general2}' using the pulsar's ephemeris, such that samples of SW DM series for any known pulsar are simulated as $\mathrm{DM}^\mathrm{SW}_\mathrm{sim} \equiv {y} {S}_{\!\theta}$. Fig.~\ref{fig: example_DMSW} shows an example of two simulated SW DM series obtained in this way. 
Note that because our method allows the spherical SW model to vary in time, the resulting simulated DM series is not strictly symmetric with respect to the solar conjunction peaks. The sampled series ${y}$ is not a discrete set of amplitudes for each solar conjunction, but rather a continuous function that varies with each observation time in a way informed by real data.

\section{Results \& Discussion} \label{sec: resultsdiscuss}

In this section, we present and discuss the results of our tests of the pipeline and of the sensitivity of available data to SW variations. In all cases, we use the software package \textsc{run\_enterprise} to simultaneously fit for the standard pulsar deterministic parameters, as well as achromatic and DM red noise with power-law priors. The deterministic parameters are marginalised over, and recovered later if necessary using least-squares fitting in \textsc{tempo2}; the hyperparameters for the red-noise power-law priors ($\log_{10}(A_\mathrm{red})$, $\gamma_\mathrm{red}$, $\log_{10}(A_\mathrm{DM})$, $\gamma_\mathrm{DM}$), and the square-root of the variance in SW yearly amplitudes ($\sigma_\mathrm{\!sw}^{}$) are sampled using a Markov Chain Monte Carlo technique through the \textsc{python} package \textsc{emcee} \citep{Foreman-Mackey2013}. In this context, we refer to `time-invariant' SW fitting as only fitting for an overall mean amplitude $N_0$ (i.e. `NE\_SW' in \textsc{tempo2}) of a spherical SW model. This is in contrast to the model described in Section~\ref{sec: swfitting}, in which we fit for a `time-variable' spherical SW. In practice, this time-variable spherical SW is described by yearly amplitude variations $\bm{\delta\!n}$ away from an overall mean amplitude $N_0$, and we fit for the $\bm{\delta\!n}$ elements, as well as for $N_0$.

\subsection{Testing the solar wind fitting code}

To initially probe the capabilities of our pipeline, we simulate a $20$-yr long set of uniform, high-cadence (one observation every 10 days) ToAs of PSR J0034$-$0534. This testing dataset includes simultaneous observations of three common frequency bands, centred at $300\,\mathrm{MHz}$, $1440\,\mathrm{MHz}$ and $2400\,\mathrm{MHz}$, with an rms of $1\,\upmu\mathrm{s}$ on each ToA. While these data are `optimistic' compared with typical pulsar datasets, they provide a useful first test to gain insight into the performance of the SW fitting pipeline. To create the simulated dataset, we start from idealised ToAs (of zero residuals), and `inject' white noise variations at the $1\text{-}\upmu\mathrm{s}$ level, Gaussian-process achromatic red noise and DM noise with known power-law priors informed by typical real levels of noise seen in the observations, as well as a realistic time-variable SW, using the method described in Section~\ref{sec: simSW}. Multiple realisations of the noise and SW Gaussian-process samples are used for robustness and repeat-measurement checks on the results of the pipeline.

In this section, we aim to explore the proficiency of our pipeline. Firstly, the injected signals are compared to the recovered ones using either the time-invariant or the time-variable SW method. The comparison of the results of the two fitting methods is discussed in Section~\ref{sec: testing_compS01}. Secondly, we use the measurements of 2000 SW amplitudes in order to validate the size of the uncertainties produced by our pipeline; this is described in Section~\ref{sec: testing_uncertainties}.

\subsubsection{Comparison between time-invariant and time-variable fitting} \label{sec: testing_compS01}

One of the benefits of using simulations is that the quantity of red noise in the pulsar ToAs is known, and we can therefore compare the properties of the recovered signals to those of the injected ones. Here, we run both the time-invariant and the time-variable fitting pipelines on the same set of simulated ToAs, which include both achromatic and DM red noise.

In general, we expect the recovered DM power-law to be affected by whether the SW influence is well modelled. Specifically, if only a time-invariant mean amplitude is used to model the SW, any year-by-year variation is likely to be absorbed by the DM noise model instead. This leads to excess power in the measured DM noise at high Fourier frequencies (at $1\,\mathrm{yr}^{-1}$ and further harmonics), which in turn causes the power-law to appear flatter than the real power-law process. Therefore, unmodelled SW influences may bias the interpretation of the DM spectrum. Furthermore, this could also bias the estimation of other timing parameters, and especially their uncertainties, as the DM power-law would not accurately represent the actual noise present in the data.
% ; this is particularly relevant for timing parameters describing periodic events.

\begin{figure}
    \centering
    \includegraphics[width=\linewidth]{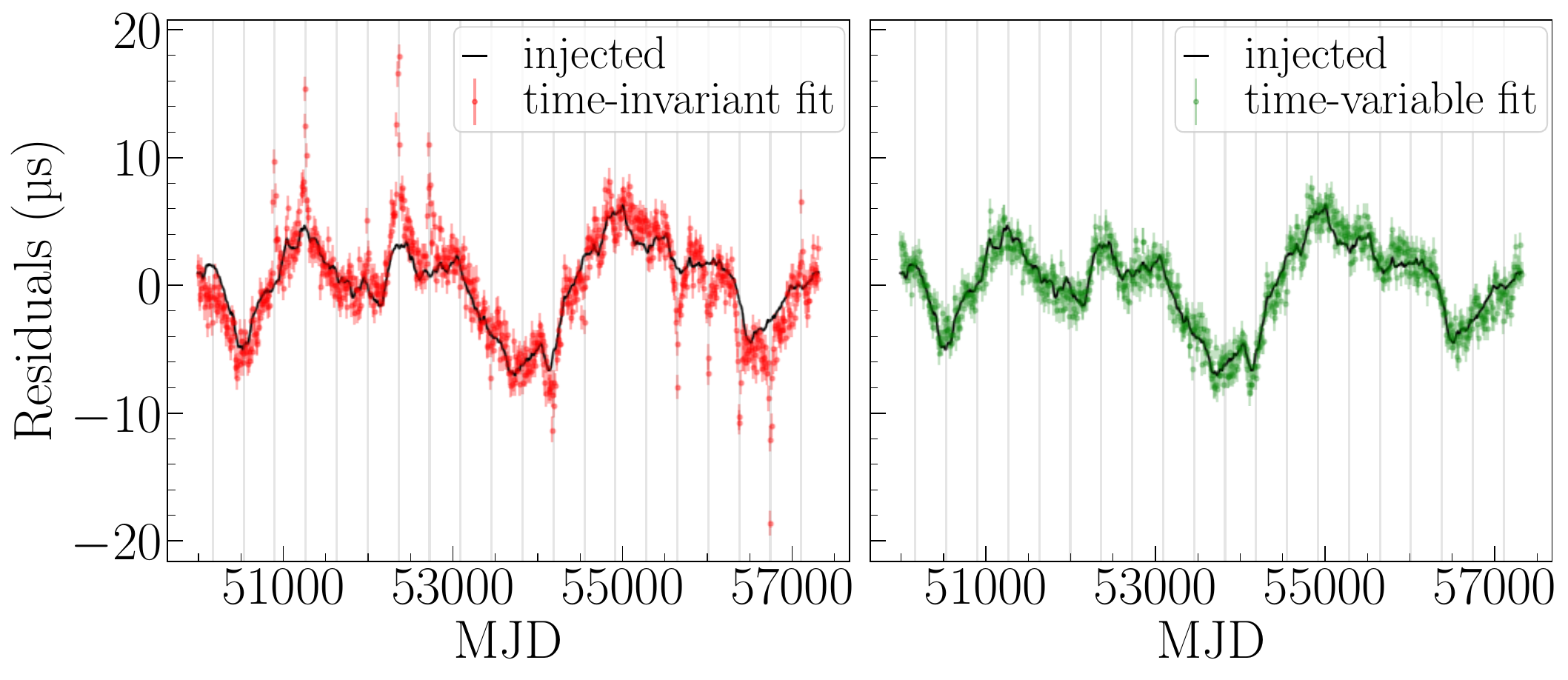}
    \caption{A single realisation of simulated residuals for the 300 MHz band, after subtracting the fitted SW variations and achromatic red noise when using the time-invariant (left) and the time-variable (right) SW amplitude model. The solid black line shows the delays at 300\,MHz for the injected IISM DM variations. Solar conjunction times are shown by the vertical gray lines.
    %The residuals shown therefore represent the fitted DM influence (without the SW) for both models. The black continuous line represents the injected DM Gaussian-process power-law time series. Only the data at an observing frequency of 300\,MHz are plotted, for clarity. Both models recover the general trend well, while the DM influence in the time-invariant fit has clearly absorbed some of the SW in the data, as can be seen by the yearly spaced peaks in the time series.
    }
    \label{fig: unifsim_residuals_justDM}
\end{figure}

To investigate the performance of our pipeline compared to the time-invariant fitting in this context, we inspect the low-frequency (300 MHz) residuals after subtracting the maximum-likelihood SW contribution and fitted achromatic noise.
Fig.~\ref{fig: unifsim_residuals_justDM} shows these residuals, overlaid with the injected IISM DM variations. 
When fitting only for a time-invariant SW, we can see additional `spikes' in the residuals around the solar conjunction times, a result of absorbing unmodelled SW into the DM variations, whilst the time-variable SW fitting is fully consistent with the injected IISM contribution.

\begin{figure}
    \centering
    \includegraphics[width=\linewidth]{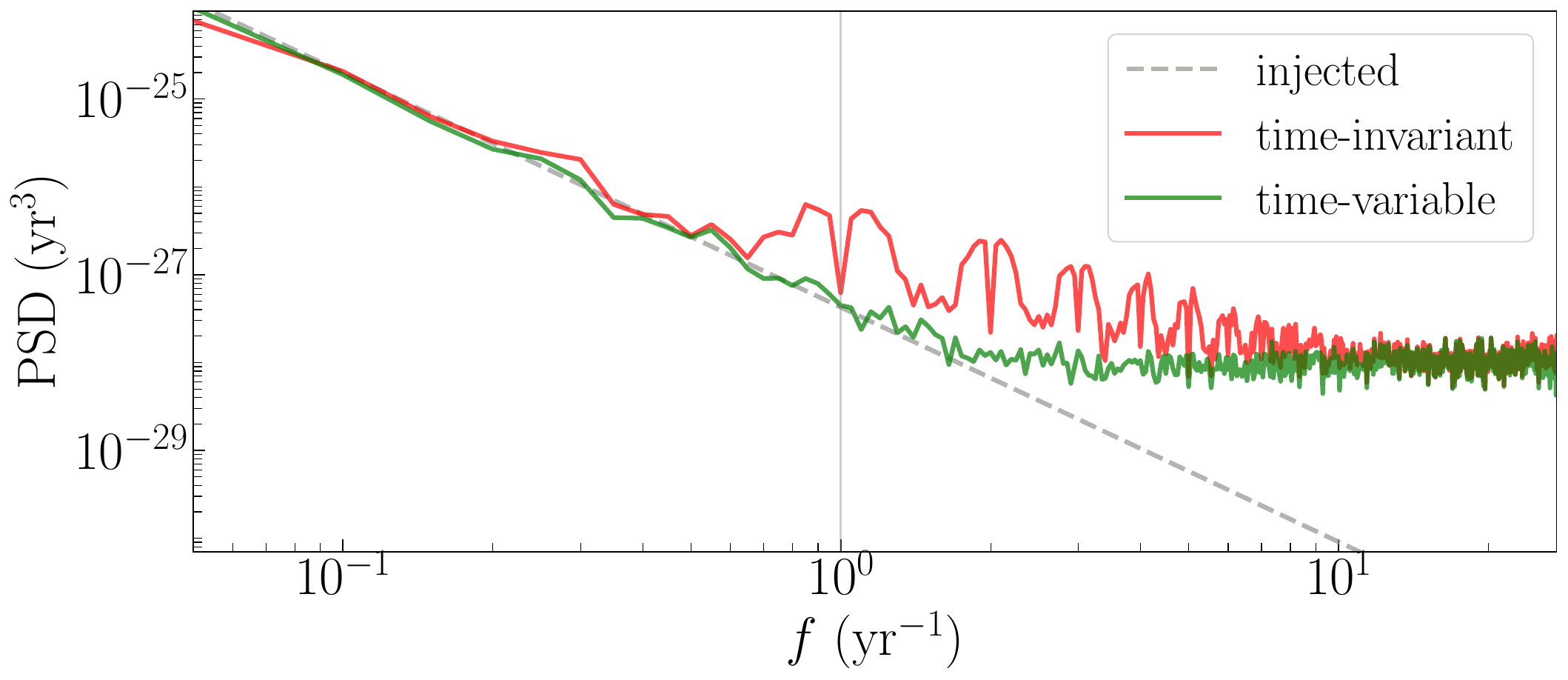}
    \caption{The power spectral density (PSD) against Fourier frequency of the recovered DM is shown, averaged over 15 realisations of our simulated data, after fitting for a time-invariant (red) and time-variable (green) SW. The grey dashed line represents the power-law prior of the injected DM Gaussian-process signal.}
    \label{fig: unifsim_DM_PSD}
\end{figure}

The average power spectral densities of these recovered DM influences, as well as the injected power-law prior, are shown in Fig.~\ref{fig: unifsim_DM_PSD}. As expected, in the time-invariant SW fit, the absorbed SW influence creates additional power at high frequencies, and therefore flattens the power-law shape compared to the injected prior; while the time-variable fit recovers the injected power-law well. The shape of the average power spectral density in the time-invariant fit shows the effect of the SW 1/yr-frequency and its harmonics, modulated by subtracting a mean-amplitude SW, which creates the observed dips of smaller width.

\begin{figure}
    \centering
    \includegraphics[width=\columnwidth]{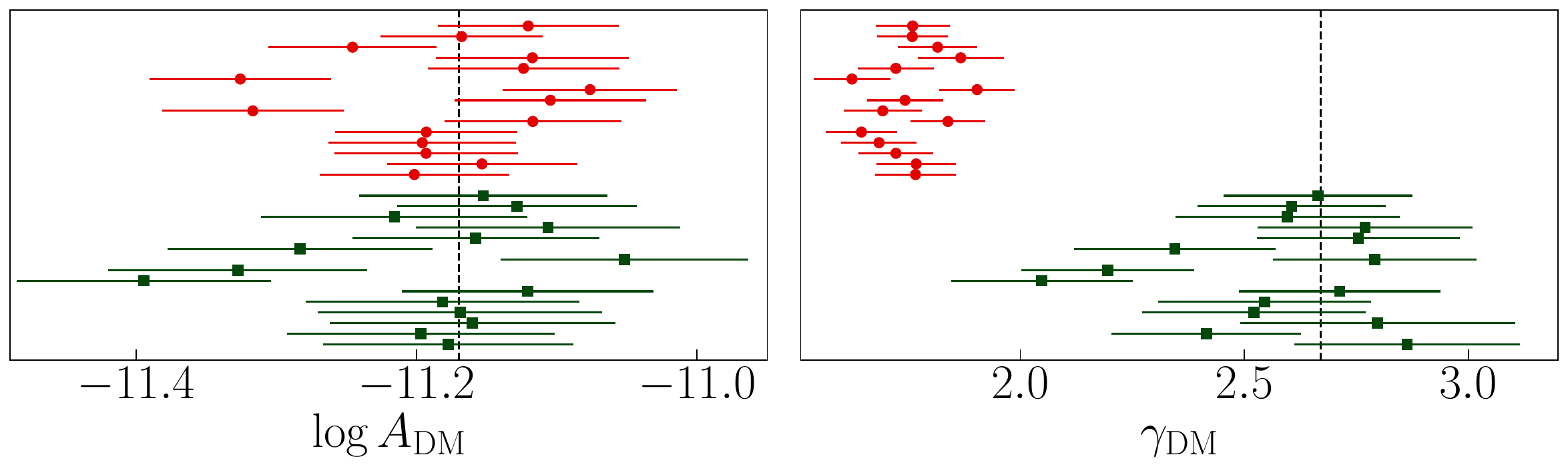}\\
    \includegraphics[width=\columnwidth]{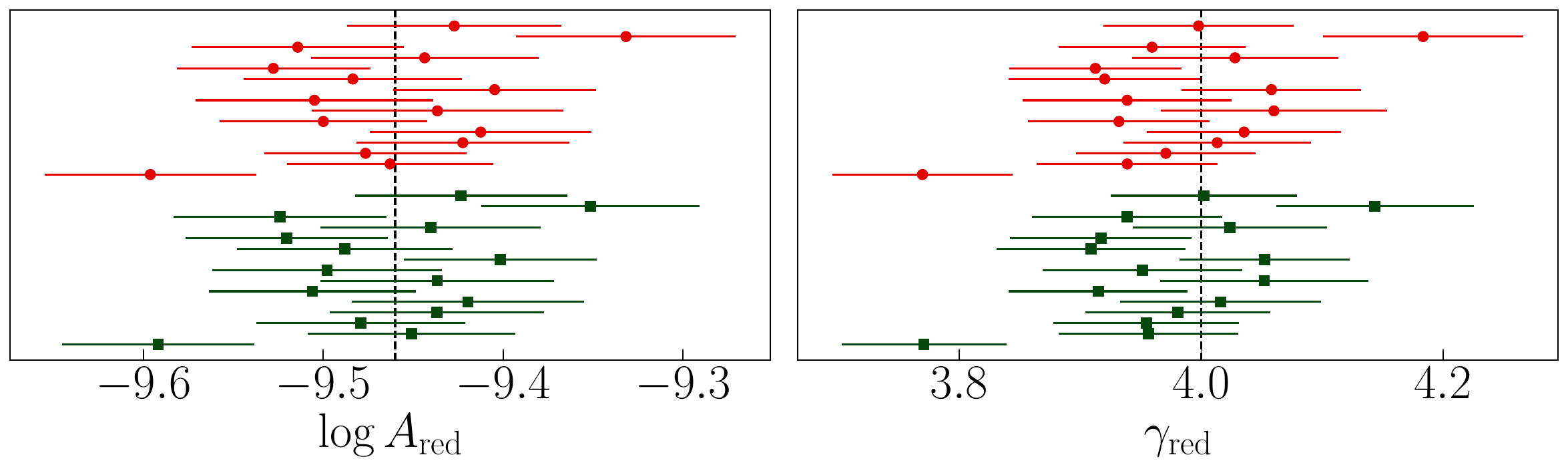}
    \caption{Measurements obtained from 15 realisations of the uniform simulation as described in Section~\ref{sec: testing_compS01} are shown. Specifically, these plots show the power-law log-amplitude ($\log{A}$) and slope ($\gamma$) for the DM (top plots) and achromatic red noise (bottom plots). The error bars represent one standard deviation. The vertical dashed lines show the `true' values used in the simulations. In each plot, 15 realisations of the simulation are shown for each of the two fitting modes: the top red circle points represent the measurements obtained from the time-invariant SW fit; while the bottom green squares correspond to the time-variable SW fitting as described in this work.}
    \label{fig: J0034_DMRed_param}
\end{figure}

\begin{figure}
    \centering
    \includegraphics[width=\columnwidth]{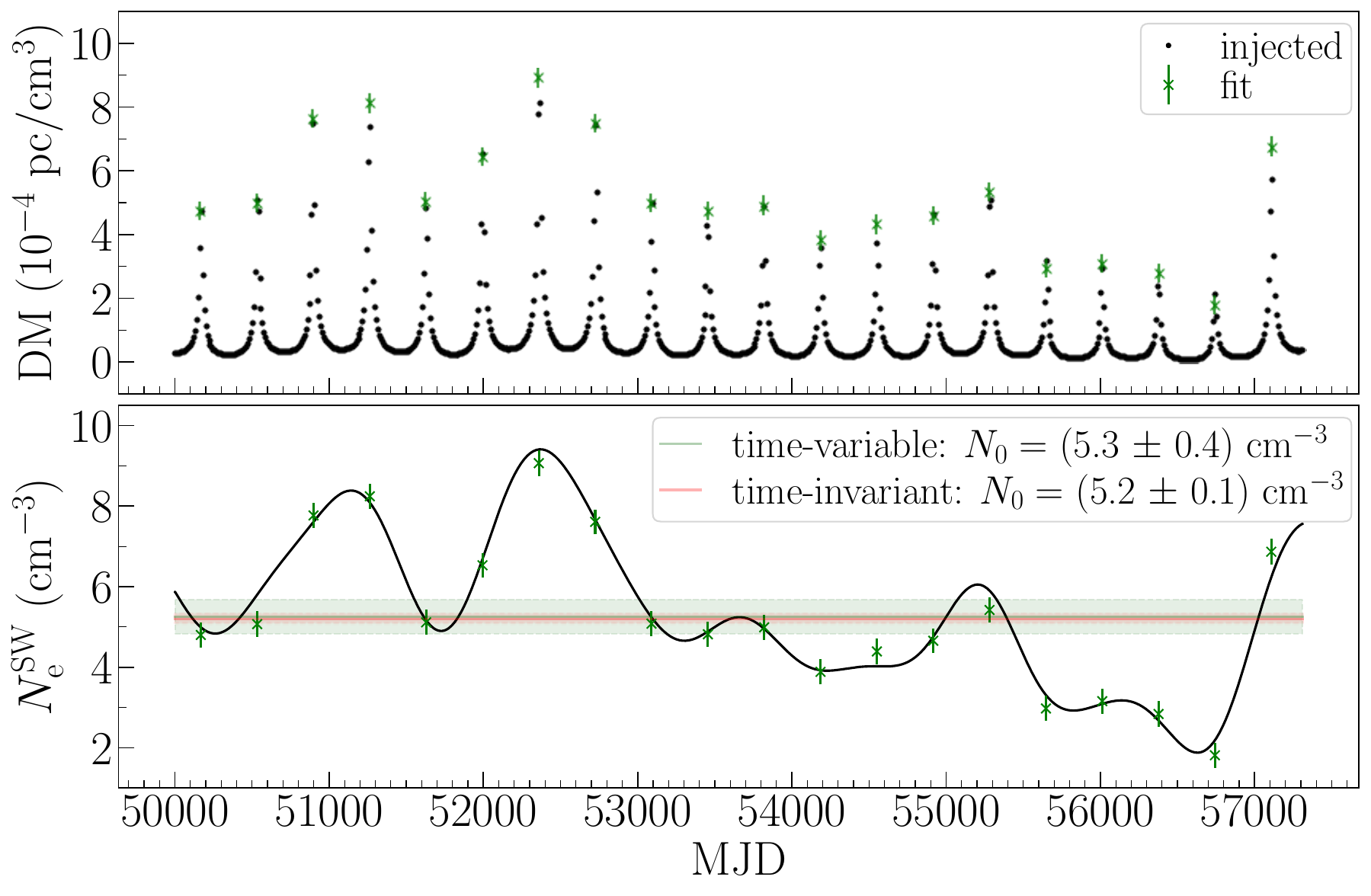}
    \caption{The injected and recovered SW contributions are compared. The top plot represents the DM series, where the black points show the injected signal, as well as illustrate the cadence of the ToAs; while the green data points are the recovered yearly amplitudes. The bottom plot represents the above DM divided by the geometrical pulsar factor $S_{\!\theta}$. The continuous horizontal lines show the mean SW amplitude ($N_0$) fitted in each of the time-variable (green) and time-invariant (red) model; the shaded region around these lines represent the area within one standard deviation.}
    \label{fig: unifsim_SWresults}
\end{figure}

The injected and recovered red noise hyperparameters for 15 realisations of the simulation are shown in Fig.~\ref{fig: J0034_DMRed_param}, as defined in the power-law in Eq.~\ref{eq: psd}. The achromatic red noise hyperparameters are in agreement with the injected values and equally well recovered in both cases; this may be expected, as the SW influence is an intrinsically chromatic effect. The time-variable SW pipeline also recovers the DM power-law well, though this is not the case for the time-invariant SW model. Notably the time-invariant model leads to a spectral exponent that is significantly flatter, with the recovered slope being of order 10-sigma smaller than the `true' value. This appears to be a systematic rather than statistical effect, i.e. not dependent on the specific Gaussian-process sample used in the simulation, but rather due to the fitting model.
\iffalse
\begin{table}
	\centering
	\caption{The injected and recovered red noise power-law hyperparameters of the simple uniform-cadence simulations. The subscript `red' denotes the achromatic red noise, while `DM' refers to the IISM DM noise. The numbers in the brackets represent one standard deviation in the last digit shown.}
	\label{tab: recovRNDM}
	\begin{tabular}{cD{.}{.}{3}D{.}{.}{6}D{.}{.}{6}} % four columns, alignment for each
		\hline
		parameter & \multicolumn{1}{c}{injected} & \multicolumn{1}{c}{time-invariant} & \multicolumn{1}{c}{time-variable}\\
		name & \multicolumn{1}{c}{value} & \multicolumn{1}{c}{fit} & \multicolumn{1}{c}{fit}\\
        \hline
		\hline
        $\log_{10} A_\mathrm{red}$ & -9.46 & -9.46(6) & -9.45(6)\\
        $\gamma_\mathrm{red}$ & 4.00 & 3.94(8) & 3.96(8)\\ 
        $\log_{10} A_\mathrm{DM}$ & -11.17 & -11.13(7) & -11.18(10)\\
        $\gamma_\mathrm{DM}$ & 2.67 & 1.76(8) & 2.45(22)\\
		\hline
	\end{tabular}
\end{table}
\fi
The recovered SW model from each fitting method can also be compared to the injected SW influence. Fig.~\ref{fig: unifsim_SWresults} shows this for the DM series (top panel), as well as the spherical SW amplitude $N_\mathrm{e}^\mathrm{SW}$ (bottom panel). 
The modelled solar-conjunction amplitudes estimated by our pipeline follow the injected SW well, being within 2-sigma of all injected values. The time-invariant fit (shown only in the bottom panel of Fig.~\ref{fig: unifsim_SWresults}) only models the SW influence with a constant-in-time amplitude, and the uncertainty in the mean amplitude recovered is too small to account for the yearly variation.

\begin{figure}
    \centering
    \includegraphics[width=\linewidth]{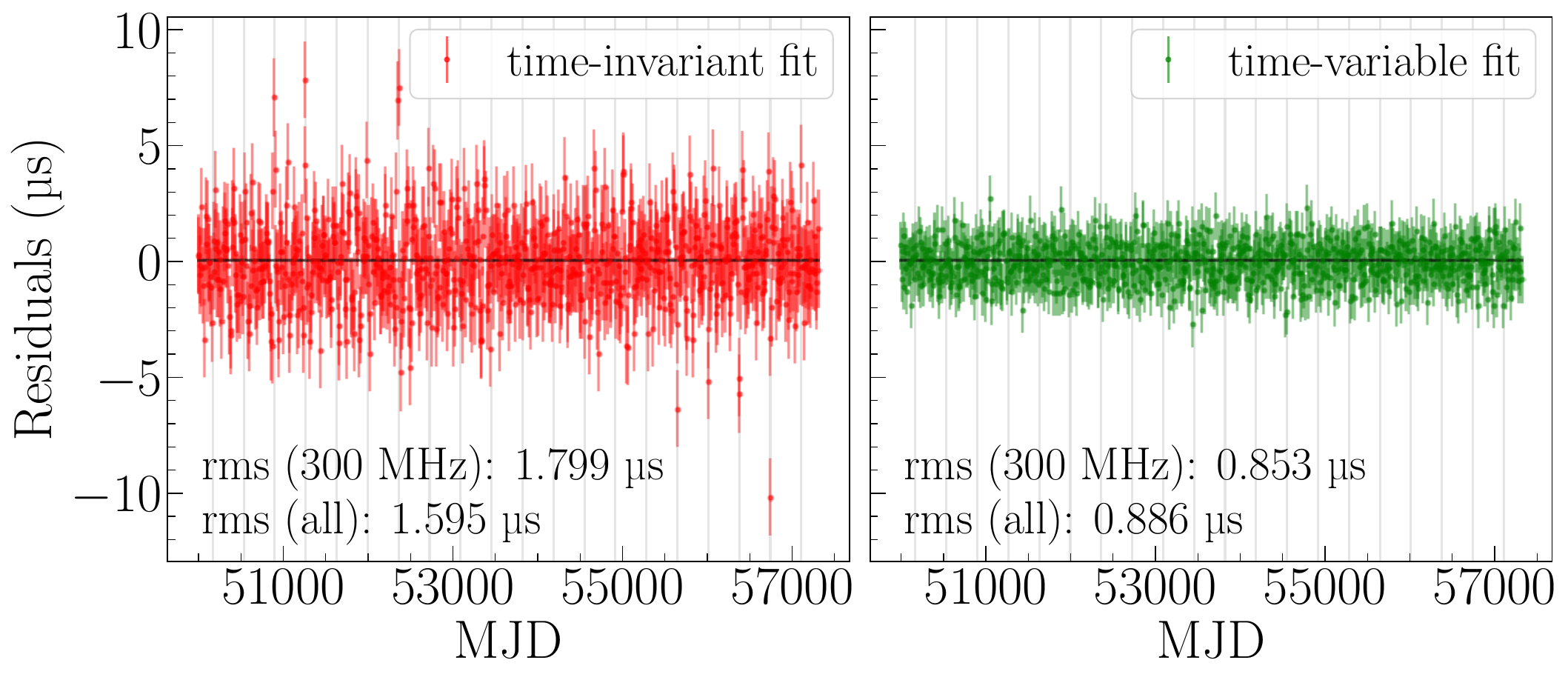}
    \caption{Post-fit residuals after subtracting the entire timing model, including achromatic and DM red noise, are shown for both the time-invariant (left) and time-variable (right) SW amplitude model. Only the data at the 300\,MHz observing frequency is shown for clarity, as it shows the largest difference between the two models. The root-mean-square (rms) values for both the 300\,MHz and the full data (`all') are also presented.}
    \label{fig: unifsim_rms}
\end{figure}

We also consider the timing residuals after removing all the modelled signals, including the achromatic red noise and DM noise; these are shown in Fig.~\ref{fig: unifsim_rms}. Ideally these fully subtracted residuals should be as close to pure uncorrelated noise as possible. The size of this left-over noise, which can be characterised by its root-mean-square (rms), is also relevant in general when searching for signals in the data ---such as, for example, the search for pulsar binary companions, or a GWB signature, for which the lower the noise the better the chance of detection. Of the two fitting modes used in this test, the yearly variable SW appears to be a better model, as strongly supported by the Bayesian evidences, which yield a natural log-Bayes factor of 160 in favour of the time-variable pipeline; the Bayesian evidences are computed using the nested-sampler \textsc{dynesty} \citep{Speagle2020}. This effect can also be seen in the fully subtracted residuals (Fig.~\ref{fig: unifsim_rms}), where the residuals obtained using the time-variable SW fitting qualitatively appear more `white', lacking the unmodelled SW `spikes'. However, note that the reduced-$\chi^2$ values of the two whitened time series are indistinguishable ($\sim\!1$), likely due to the large number of data points not at solar conjunctions. The level of white noise, quantified here by the rms, is better by $\sim\!1\,\upmu\mathrm{s}$ for the residuals obtained from the time-variable fitting than those from the time-invariant fitting.

\subsubsection{Recovered uncertainties} \label{sec: testing_uncertainties}

To check the robustness of our results and their uncertainties, 100 sets of ToAs are simulated, all with the same general properties as the dataset used in Section~\ref{sec: testing_compS01}. Each of these simulated datasets is created from different Gaussian-process realisations of the same power-law prior achromatic red noise and DM noise, as well as different samples of SW variations. The results of running these 100 independent datasets through our pipeline are consistent with the findings in Section~\ref{sec: testing_compS01} with respect to the injected values.

The 2000 solar conjunctions available in total in the 100 datasets are used to study the statistical properties of the estimated uncertainties on the SW amplitudes. Each measurement of the SW amplitude is normalised with respect to the injected value and the associated uncertainty, such that the equivalent `standardised variable' is computed. The population of these standardised variables is expected to be normally distributed, with a mean of zero and a standard deviation of one, if the distribution of the measured variable is Gaussian. In the case of a set of uncorrelated measurements, $\{X_i\}$, assumed to be Gaussian distributed, the standardised variable is simply defined as $(X_i - \mu)/\sigma$ for each measurement, where $\mu$ and $\sigma$ are the mean and standard deviation of the Gaussian distribution. In this analysis, however, the measurements of SW amplitudes are correlated through the mean $N_0$, such that we use linear algebra to compute the standardised variable values, as follows.

If $\bm{n}$ is the column vector of the $m\!=\!2000$ measured SW amplitudes, and $\bm{n}^{\mathrm{inj}}$ is the equivalent column vector of injected SW amplitudes, then we can write
\begin{equation}
    \bm{n} = \bm{n}^{\mathrm{inj}} + \bm{e},
\end{equation}
where $\bm{e}$ is a column vector representing the random variation of $\bm{n}$ around $\bm{n}^{\mathrm{inj}}$, expected to be normally distributed with mean zero and variance according to the parameter covariance matrix ${\mathbf{C}}$ of the elements in $\bm{n}$. If ${\mathbf{C}}$ is known, the vector $\bm{e} = \bm{n} - \bm{n}^{\mathrm{inj}}$ can be `whitened' to disentangle the correlations between different amplitude measurements, and therefore to obtain the column vector of standardised variables $\bm{z}$. This approach is similar to e.g. the whitening presented in \citet{Coles2011} for general pulsar data, and we summarise the steps below. Using the Cholesky decomposition, the parameter covariance matrix is written as ${\mathbf{C}} = {\mathbf{L}} {\mathbf{L}}^{\!\!\top}$, where ${\mathbf{L}}$ is a lower triangular matrix with a real and positive diagonal. The whitening process then yields that the column vector of standardised variables can be computed as
\begin{equation}
    \label{eq: zL-1}
    \bm{z} = {\mathbf{L}}^{-1} \bm{e} = {\mathbf{L}}^{-1} (\bm{n} - \bm{n}^{\mathrm{inj}}).
\end{equation}
Recall that in our pipeline, we fit for the mean SW amplitude $N_0$ and the deviations from it at each solar conjunction, i.e. $\bm{\delta \! n} = \bm{n} - N_0 \mathrm{\textit{\textbf{1}}}$, where in this case $\mathrm{\textit{\textbf{1}}}$ is an $m$-point column vector of ones. The relevant column vector of measured SW parameters is therefore given by the ($m\!+\!1$)-point column vector $\bm{p} = (N_0, \bm{\delta\!n})^{\!\top}$, with the associated parameter covariance matrix, $\mathbf{C}_{\bm{p}}$. The parameter covariance matrix $\bm{C}_{\bm{p}}$ is a direct result of the timing analysis, and can be obtained directly from \textsc{tempo2} post-fitting. The whitening covariance matrix $\mathbf{C}$ can then be computed as 
\begin{equation}
    \label{eq: transformpn}
    {\mathbf{C}} = {\mathbf{M}} \mathbf{C}_{\bm{p}} {\mathbf{M}}^{\!\top},
\end{equation}
where ${\mathbf{M}}$ is the transformation matrix from the ($m\!+\!1$)-point column vector $\bm{p}$ to the $m$-point column vector $\bm{n}$, i.e. $\bm{n} = {\mathbf{M}} \bm{p}$, and therefore has the shape
\begin{equation}
    \label{eq: M}
    {\mathbf{M}} = 
    \begin{pmatrix}
      1 & 1 & 0 & \cdots & 0 \\
      1 & 0 & 1 & \cdots & 0 \\
      \vdots & \vdots & \vdots & \ddots & \vdots \\
      1 & 0 & 0 & \cdots & 1
    \end{pmatrix},
\end{equation}
where the first column contains exclusively `1's and the rest of the matrix has the shape of an $(m\times m)$ identity matrix.
Therefore, the standardised variable $\bm{z}$ can be computed as per Eq.~\ref{eq: zL-1} using the Cholesky decomposition of the parameter covariance matrix ${\mathbf{C}}$, which can be estimated from the output covariance matrix of the analysis, $\mathbf{C}_{\bm{p}}$, using Eq.~\ref{eq: transformpn} and the shape of the transformation matrix, ${\mathbf{M}}$, as given in Eq.~\ref{eq: M}. 

\begin{figure}
    \centering
    \includegraphics[width=\columnwidth]{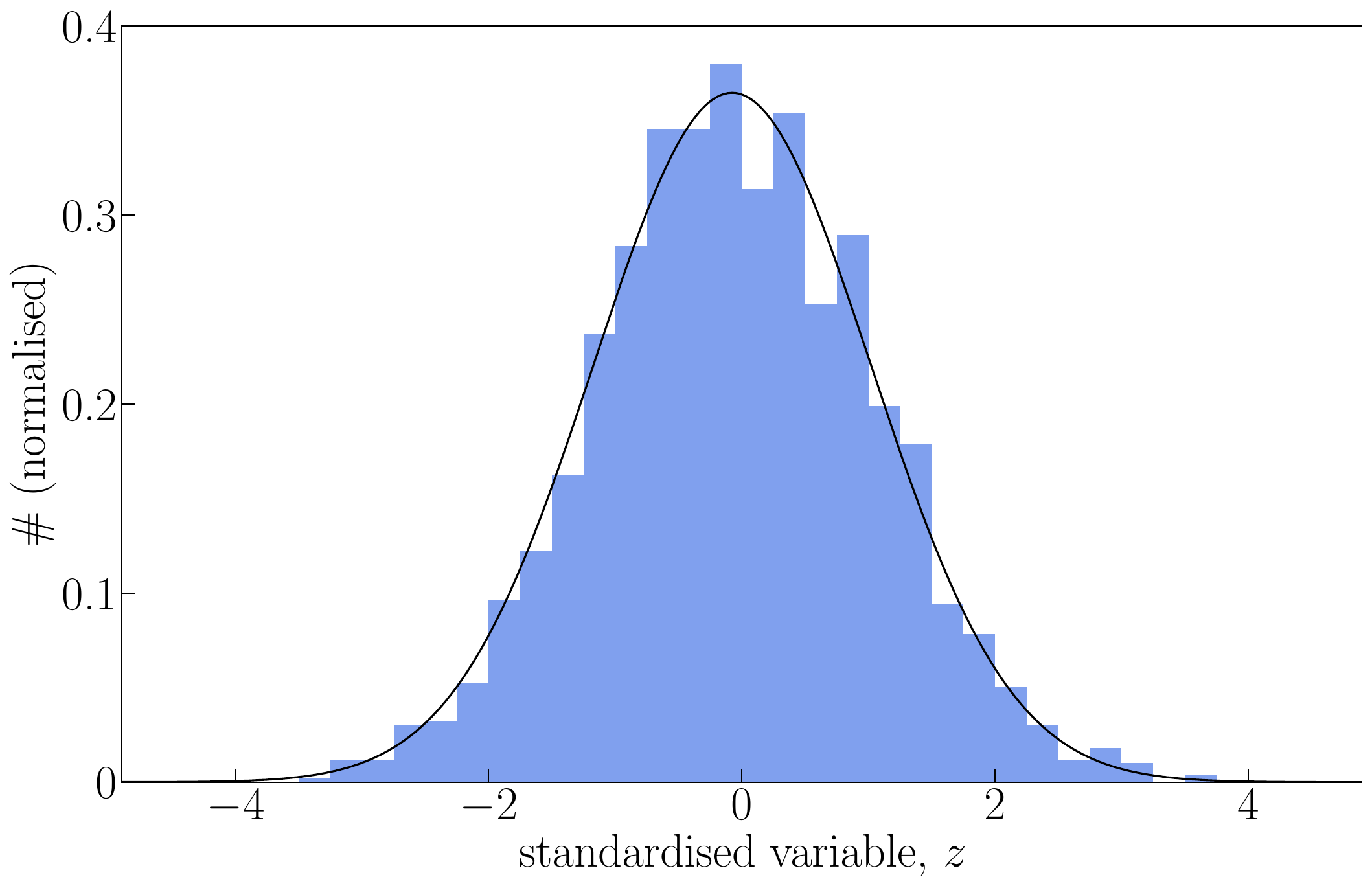}
    \caption{The histogram of the standardised variable ($z$) obtained from the 2000 recovered solar conjunction amplitude values and uncertainties. The best-fit Gaussian distribution is also shown, with a mean of $-0.05(2)$ and a standard deviation of $1.09(2)$.}
    \label{fig: unifsim_SWhist}
\end{figure}

Fig.~\ref{fig: unifsim_SWhist} shows a histogram of the (normalised) distribution of the measurements of $\bm{z}$ for the 2000 solar-conjunctions amplitudes fitted with our pipeline. 
The distribution of this rescaled uncertainty is well described by a Normal distribution, with an Anderson-Darling test equivalent $p$-value of 0.55 \citep[e.g.][]{DAgostino1986}. The Normal distribution has mean of $-0.05(2)$ and a standard deviation of $1.09(2)$, not meaningfully distinguishable from a unit-variance zero-mean Normal distribution; we therefore conclude that our uncertainties are well estimated within our assumptions.

\subsection{Solar wind fitting in PTA-type simulations}

In this section, we aim to establish whether PTA-like data are sensitive to yearly variations in SW away from the mean amplitude. For this purpose, EPTA+InPTA data of PSR J1022$+$1001 are used as a basis for simulated ToAs which are then analysed through our pipeline. We use the real cadence and uncertainties of this dataset to simulate current observations (Section~\ref{sec: currentEPTA}), as well as median properties of relevant telescopes to approximate future observations of the EPTA+InPTA (Section~\ref{sec: futureEPTA}). 

\subsubsection{Current EPTA + InPTA observations} \label{sec: currentEPTA}

To simulate a set of ToAs with properties characteristic to the EPTA+InPTA dataset, we start from the real data of PSR J1022$+$1001, which shows the highest SW influence of the 25 pulsars included in the EPTA DR2 (as this pulsar has the closest approach to the Sun, having an ecliptic latitude of $-0.06^\circ$). Idealised ToAs (of zero residuals) are created such that the observing cadence, frequencies, and error bars of the real data are preserved into the simulated ToAs. As before, realistic levels of noise and SW variations, based on observations, are added to these idealised ToAs. 
%The achromatic red noise and DM noise power-law parameter values are set to the fitted values found in the real EPTA DR2 data for this pulsar, as per \citet{EPTA2023II}.

\begin{figure}
    \centering
    \includegraphics[width=\columnwidth]{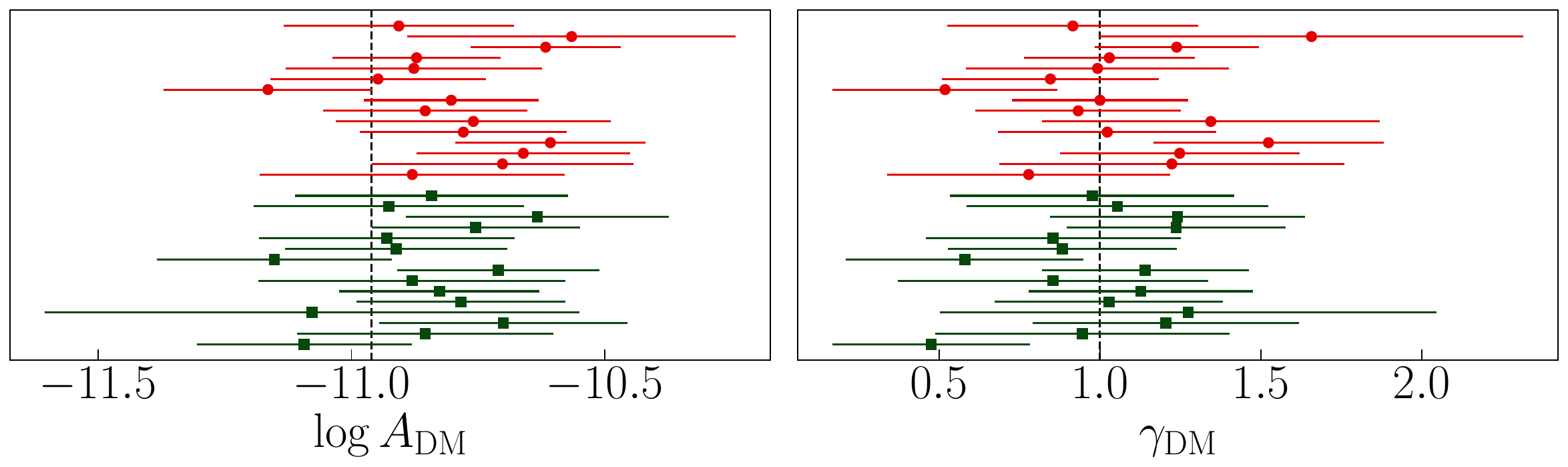}\\
    \includegraphics[width=\columnwidth]{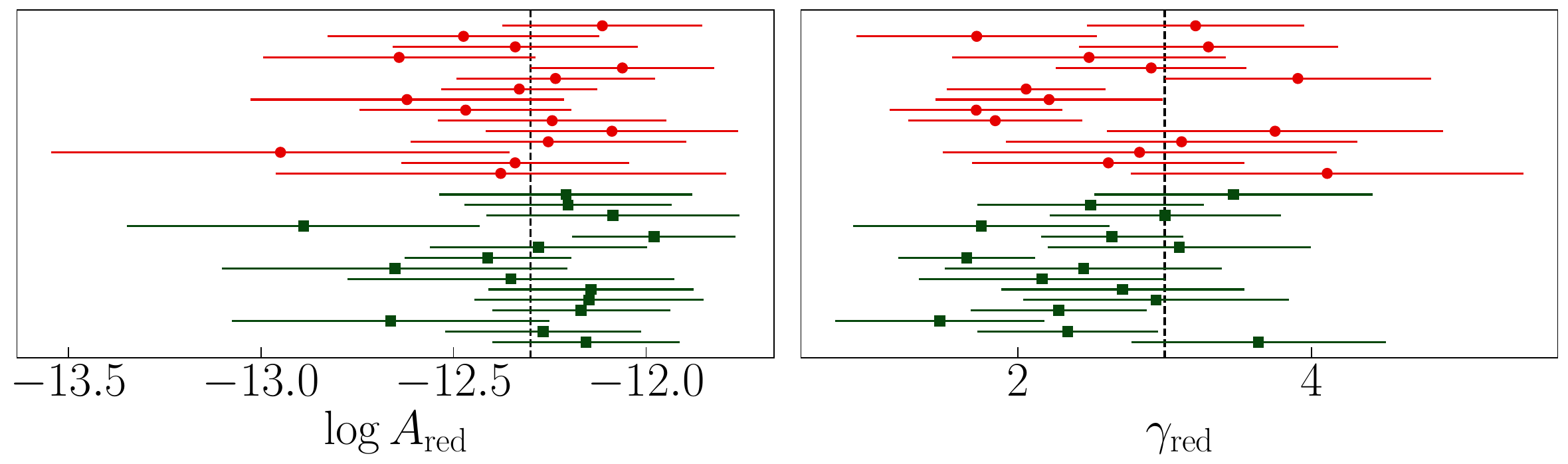}
    \caption{Measurements obtained from 15 realisations of the (EPTA+InPTA)-like simulation as described in Section~\ref{sec: currentEPTA} are shown. See the caption of Fig.~\ref{fig: J0034_DMRed_param} for details.}
    \label{fig: J1022_DMRed_param}
\end{figure}

The comparison of the measured red-noise hyperparameters with the injected values is presented in Fig.~\ref{fig: J1022_DMRed_param} for both the time-variable and the time-invariant SW fitting.
\iffalse
\begin{table}
	\centering
	\caption{The injected and recovered red noise power-law hyperparameters of the EPTA+InPTA simulations of PSR J1022$+$1001. The numbers in the brackets represent one standard deviation in the last digit shown.}
	\label{tab: EPTAlikePL}
	\begin{tabular}{cD{.}{.}{3}D{.}{.}{6}D{.}{.}{6}} 
		\hline
		parameter & \multicolumn{1}{c}{injected} & \multicolumn{1}{c}{time-invariant} & \multicolumn{1}{c}{time-variable} \\
        name & \multicolumn{1}{c}{value} & \multicolumn{1}{c}{fit} & \multicolumn{1}{c}{fit} \\
		\hline
		\hline
        $\log_{10} A_\mathrm{red}$ & -12.3 & -12.2(1) & -12.2(1)\\
        $\gamma_\mathrm{red}$ & 3.4 & 3.3(7) & 3.2(7)\\ 
        $\log_{10} A_\mathrm{DM}$ & -11.2 & -11.2(1) & -11.2(1)\\ 
        $\gamma_\mathrm{DM}$ & 0.4 & 0.1(1) & 0.1(1)\\
		\hline
	\end{tabular}
\end{table}
\fi
It can be seen that there is no significant difference between the results of the two fitting modes in this case, and that the best-fit values are mostly consistent with the `true' values within their uncertainties. However, the precision of these measurements is, as one would expect, worse than in our previous simulations of uniform cadence and low-frequency, discussed in Section~\ref{sec: testing_compS01}. 

\begin{figure}
    \centering
    \includegraphics[width=\columnwidth]{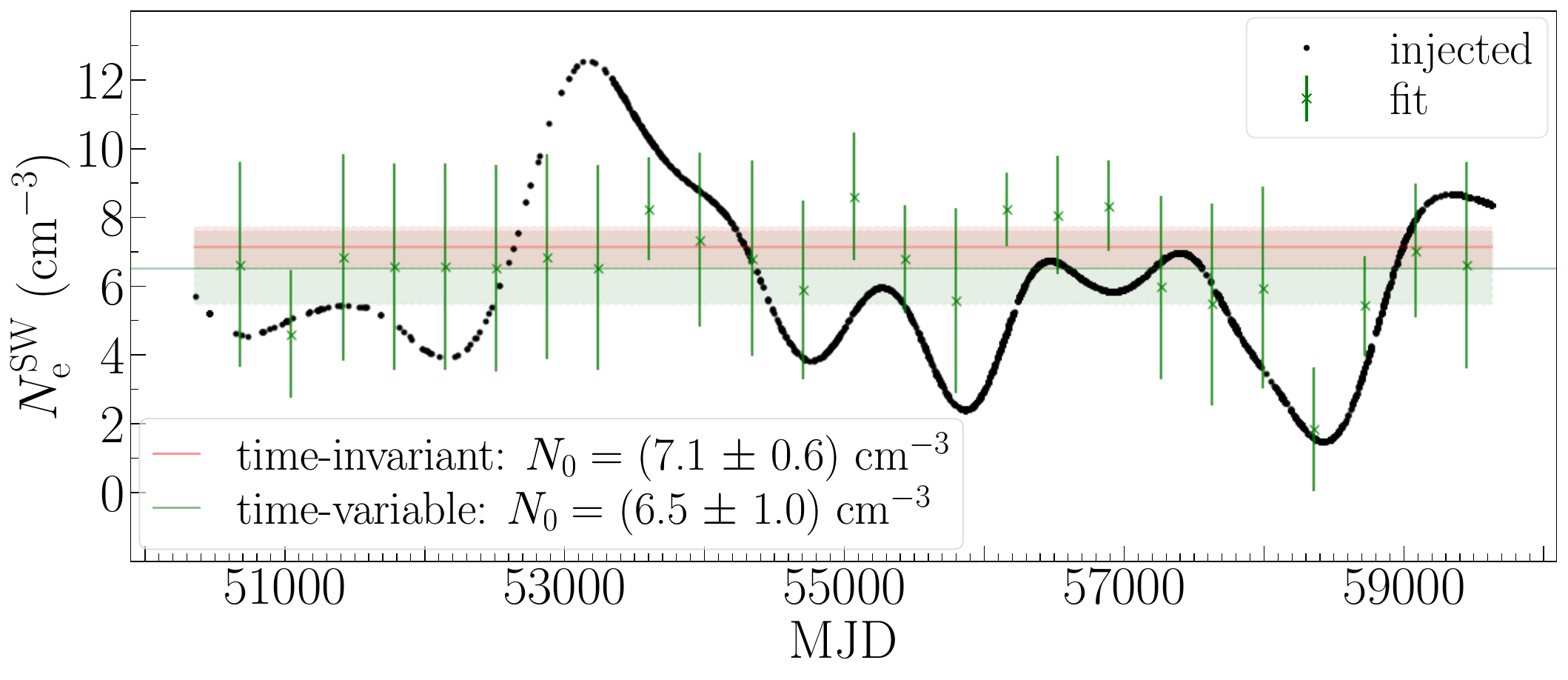}
    \caption{The injected and recovered SW contributions for the EPTA+InPTA type simulations are compared. The black points show the injected signal, as well as the cadence of the ToAs; while the green data points are the recovered yearly amplitudes. The continuous horizontal lines show the mean SW amplitude ($N_0$) fitted in each of the time-variable (green) and time-invariant (red) model; the shaded region around these lines represent the area within one standard deviation.}
    \label{fig: EPTArealsim_SWresults}
\end{figure}

Fig.~\ref{fig: EPTArealsim_SWresults} shows the fitted SW amplitudes through our pipeline, compared to the injected Gaussian process sample. 
While our results are generally consistent with the injected values, the uncertainties are large enough that the results for this dataset are also consistent with a constant-in-time SW. This outcome is also confirmed through an examination of the residuals after subtracting both the time-variable fitting model and the time-invariant model, which reveal no significant difference between the properties of the two: the reduced-$\chi^2 \simeq 0.97$ in both cases; while the rms values are $1.37\,\upmu\mathrm{s}$ and $1.33\,\upmu\mathrm{s}$ for the time-invariant and time-variable fitting, respectively.

We therefore conclude that the current EPTA+InPTA dataset is not sensitive enough to measure yearly changes in the SW amplitude. However, we also note that there is no noticeable disadvantage of including this additional model in the fitting, as it adds no significant time to the computational run, nor does it increase the overall left-over white noise level. Moreover, when there is not sufficient data around a solar conjunction to measure the variation in SW amplitude, the total amplitude value defaults to the mean (as can be seen especially in the early data in Fig.~\ref{fig: EPTArealsim_SWresults}). The lack of sensitivity to measuring SW changes of even the more recent EPTA+InPTA data is indicative of the small number of observations at low frequencies ($\lesssim 1000\,\mathrm{MHz}$) ---where the SW influence is larger and therefore easier to quantify -, as well as of the relatively small fractional bandwidths of the observations. However, this is expected to change in the future, as more uGMRT (InPTA) data, which includes high-quality simultaneous observations at low frequencies ($\sim400\,\mathrm{MHz}$), is combined into the EPTA dataset. Note that the current data combination, as used in the simulations in this section, includes some InPTA data, particularly around the solar conjunction around MJD\,58360; these InPTA observations are indeed seen to have a plausibly beneficial effect on SW amplitude fitting for that year. Furthermore, the upcoming inclusion of the high-cadence and very low-frequency LOFAR data into the EPTA dataset will unlock next-level sensitivity to the SW.

\subsubsection{Simulated `future' EPTA+InPTA observations}
\label{sec: futureEPTA}
With the results of the realistic (EPTA+InPTA)-like simulations in mind (as presented in Section~\ref{sec: currentEPTA}), we aim to investigate the sensitivity of this dataset after 10 years of observations at the current observing setup. For this purpose, we create sets of simulations based on the median of recent properties of the LEAP observing system (which is the most sensitive L-band `observatory' in the EPTA dataset), and on the median of recent properties of uGMRT observations. The properties of these simulated observations are summarised in Table~\ref{tab: futuresim}, showing the number of frequency sub-bands for each observing system, the total frequency range, the (uniform) observing cadence, and the (uniform) rms characterising the system white noise.
\begin{table}
	\centering
	\caption{The properties of the uniform-cadence simulations based on median real-data values.}
	\label{tab: futuresim}
	\begin{tabular}{ccD{-}{-}{4}cr} % four columns, alignment for each
		\hline
		observing & \# freq. & \multicolumn{1}{c}{freq. range} & cadence & rms\\
		system & bands & \multicolumn{1}{c}{[MHz]} & [d] & [$\upmu$s]\\
		\hline
		\hline
		LEAP & 1 & \multicolumn{1}{c}{1440} & 30 & 0.25\\
		uGMRT (B3) & 32 & 300-500 & 12 & 7\\
		uGMRT (B5) & 4 & 1260-1460 & 12 & 22\\
		\hline
	\end{tabular}
\end{table}
Note that the observations with the two uGMRT frequency bands (B3 and B5) are simultaneous, which is particularly useful for studying chromatic effects. Uniform-cadence observations are used in this case for simplicity; while the sensitivity of any pulsar data to the SW indeed should depend on the density of observations close to its solar conjunction, a detailed study on how this cadence affects measurements and on determining the best observing strategy is beyond the scope of the current work, and is left as a future investigation.

\begin{figure}
    \centering
    \includegraphics[width=\linewidth]{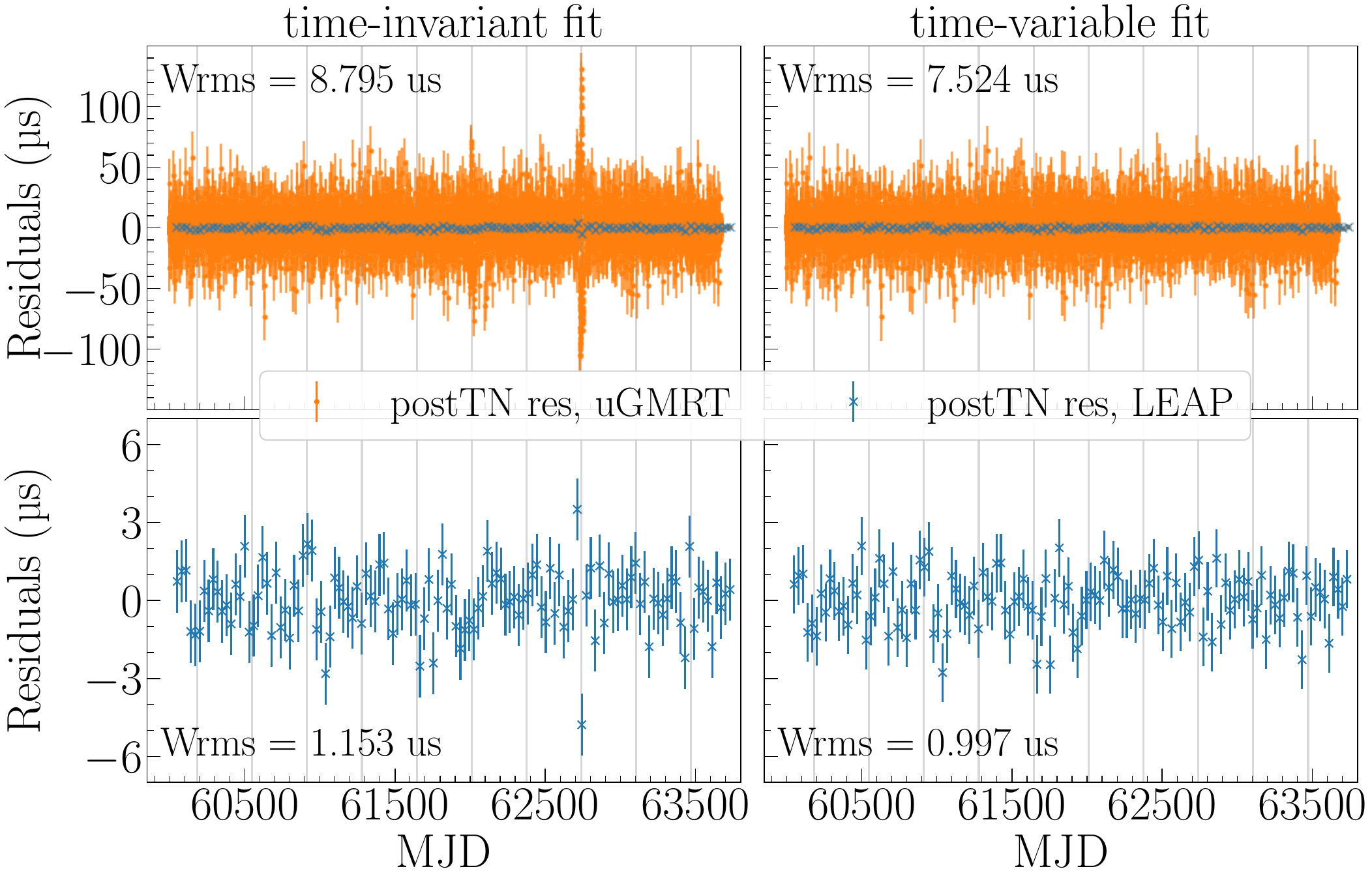}
    \caption{All plots show the post-fit residuals after removing the fitted achromatic and DM Gaussian-processes red noise. The plots on the left correspond to the results after using the time-invariant SW amplitude fit, while those on the right correspond to the time-variable fit. The top plots show the full simulated data, with the uGMRT-type data in orange and the LEAP-like data in blue. The bottom plots show the same information, but zoomed in and just for the LEAP data. The weighted root-mean-square (Wrms) is given in each case for all the data shown in the plot.}
    \label{fig: J1022_futsim_res}
\end{figure}

\begin{figure} 
    \centering
    \includegraphics[width=\linewidth]{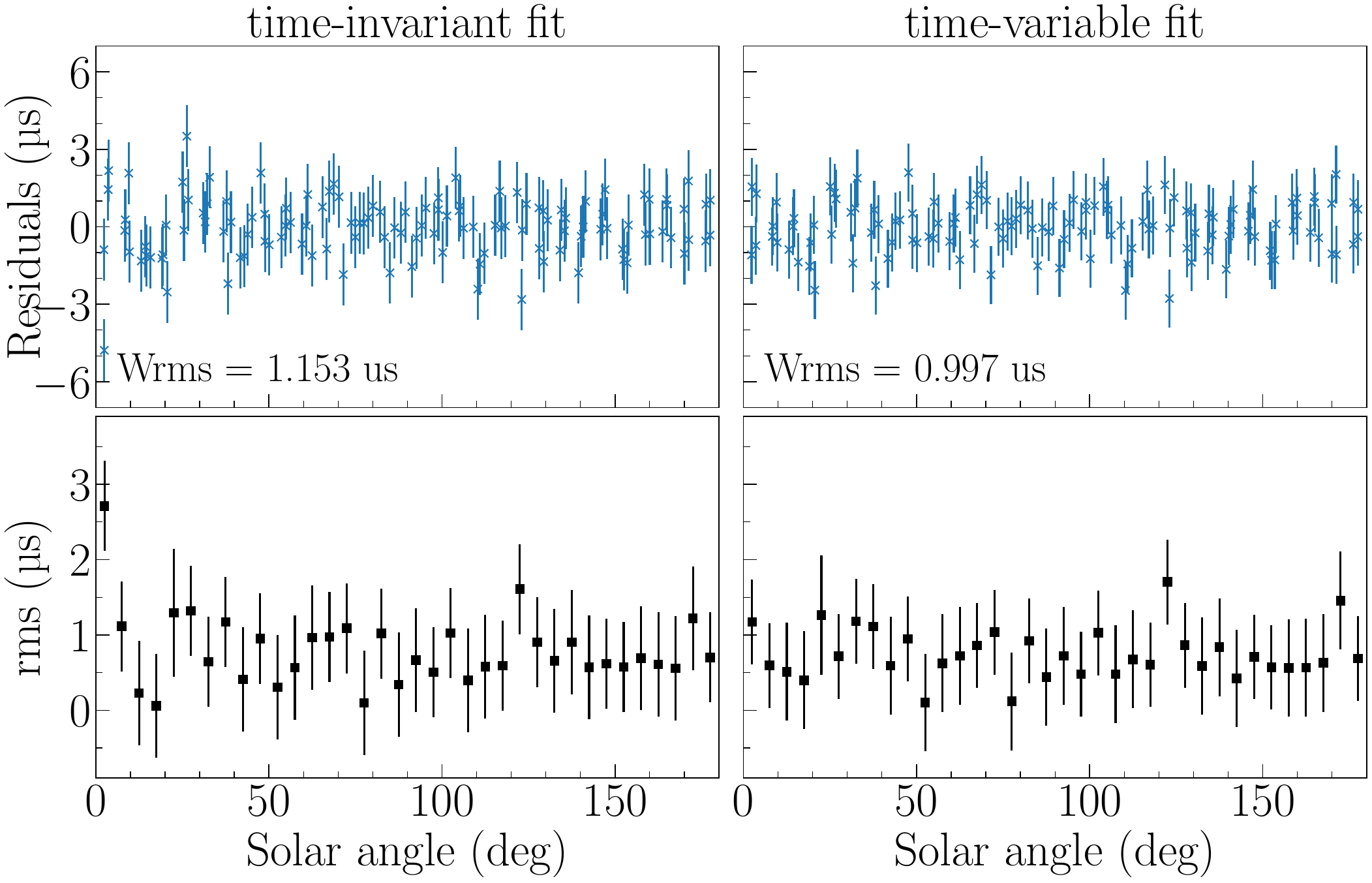}
    \caption{The data shown in the top plots is similar to the LEAP residuals shown in the bottom plots of Fig.~\ref{fig: J1022_futsim_res}, but the residuals are plotted against solar angle, i.e. the sky distance between the pulsar line-of-sight and the Sun. Again, the plots on the left correspond to the time-invariant SW amplitude fit, while the plots on the right show the results of the time-variable fit. The top plots show the residuals for every individual data point, while the bottom plots show the rms of these residuals in bins of $5^\circ$ solar angle. Of particular interest are the points at low solar angle, i.e. where the pulsar is close to solar conjunction and the SW influence is the largest.}
    \label{fig: J1022_futsim_rms}
\end{figure}

Here, we consider the improvement of the noise budget of the observations when using the time-variable amplitude SW fitting instead of the time-invariant SW fitting. Fig.~\ref{fig: J1022_futsim_res} shows the residuals in both cases, after subtracting all the fitted signals, including the red noise processes. 
The time-variable fit generally performs better in whitening the pulsar ToAs. Qualitatively, there are obvious signatures of the SW that are not removed in the time-invariant case, particularly near the solar conjunction close to MJD\,62740, visible both in low-frequency uGMRT data and, to a lesser extent, in L-band LEAP observations. Quantitatively, while the reduced-$\chi^2 \simeq 1$ in both cases as before, the overall left-over noise level is reduced by using our fitting compared to the time-invariant mean: for the full simulated dataset, the rms is reduced by nearly $1.3\,\upmu\mathrm{s}$; while just for the L-band LEAP data, the rms is improved by $160\,\mathrm{ns}$. The plots of Fig.~\ref{fig: J1022_futsim_rms} show the same LEAP residuals, but as a function of the solar angle of the pulsar, i.e. its angular distance to the Sun as projected on the sky, for both the time-invariant and time-variable fitting. 
These plots show that the main difference between the two fitting methods is, as expected, due mainly to those observations closest to the Sun, although even some observations at tens of degrees of solar angle are affected. 

Overall, we conclude that in the near future, as more InPTA observations will be combined in the EPTA datasets, the sensitivity to the SW in pulsar data will increase. Moreover, by fitting this data for a time-variable spherical SW, the noise budget is likely to improve, even for L-band observations. This is particularly interesting in the context of e.g. detecting the GWB by PTAs, where a decrease in noise of just a few hundred nanoseconds could be valuable in reaching the target detection significance in light of the recent GWB results \citep{EPTA2023I,EPTA2023II,EPTA2023III}.

\subsection{Real data}

We also test our pipeline on real data, of PSRs J0030$+$0451, J1022$+$1001, and J2145$-$0750, respectively. Firstly, our pipeline's SW amplitude estimates from independent datasets taken by the observing systems EPTA+InPTA, PPTA, and LOFAR are compared in Section~\ref{sec: compareNESW}. Secondly, the recovered DM time-series from the fits on LOFAR data are compared to those given in T21, which used the same (while slightly shorter) LOFAR datasets, but a different method; we are therefore able to directly compare SW results between our pipeline and another, independent method; this is discussed in Section~\ref{sec: compareLOFAR}.

\subsubsection{Comparison between results from different datasets} \label{sec: compareNESW}

\begin{figure}
    \centering
    \includegraphics[width=\linewidth]{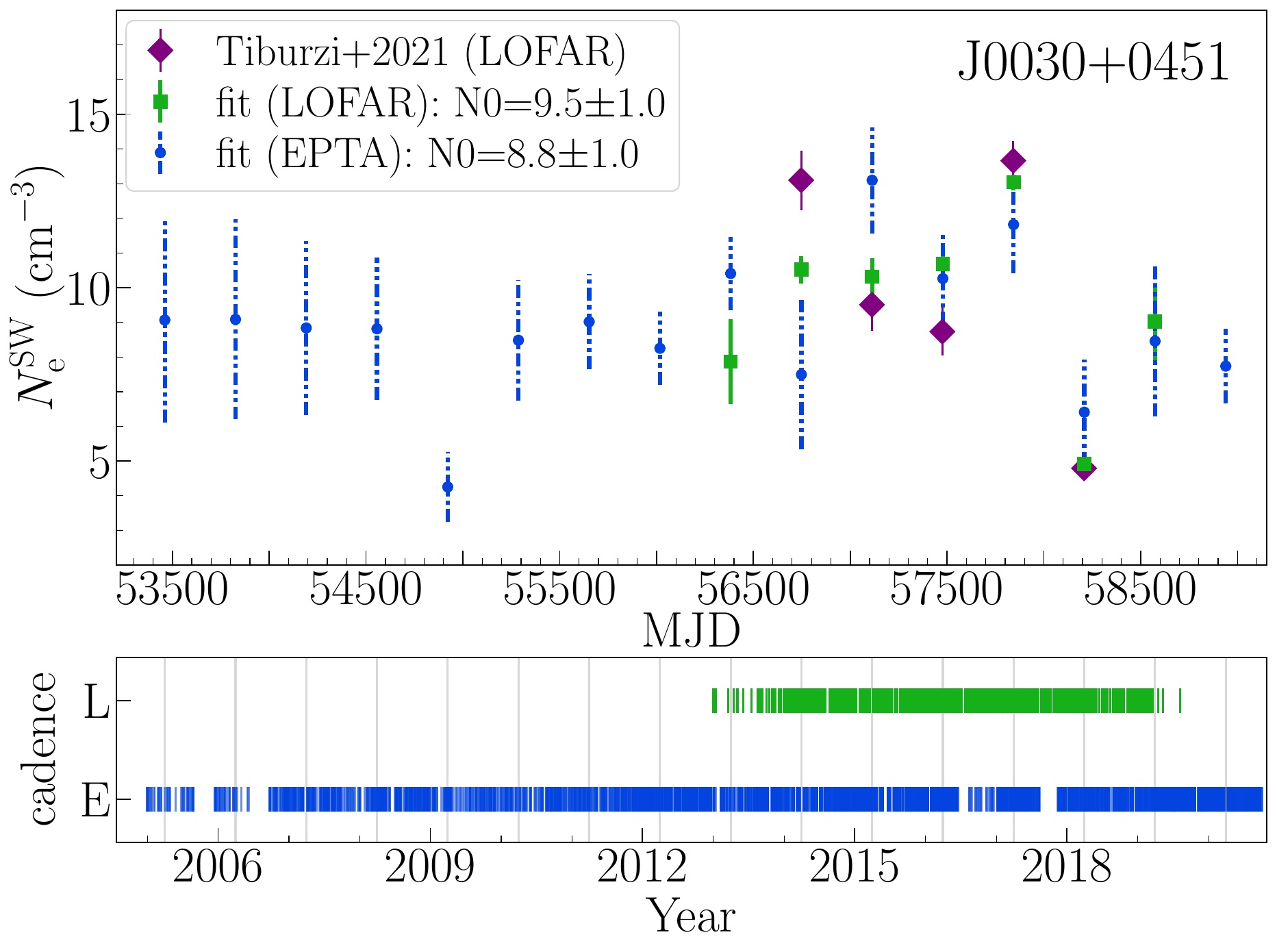}
    \caption{The yearly SW amplitudes for PSR J0030$+$0451 are shown, resulting from the EPTA (blue dots) and LOFAR (green squares) data, as well as the corresponding values presented in T21 (purple diamonds). The cadences of the observations used are also shown, where `E' stands for the EPTA data, and `L' for the LOFAR data. Note that the two x-axes are the same, but presented in both MJD and Year for convenience.}
    \label{fig: J0030_realruns}
\end{figure}

\begin{figure}
    \centering
	\includegraphics[width=\linewidth]{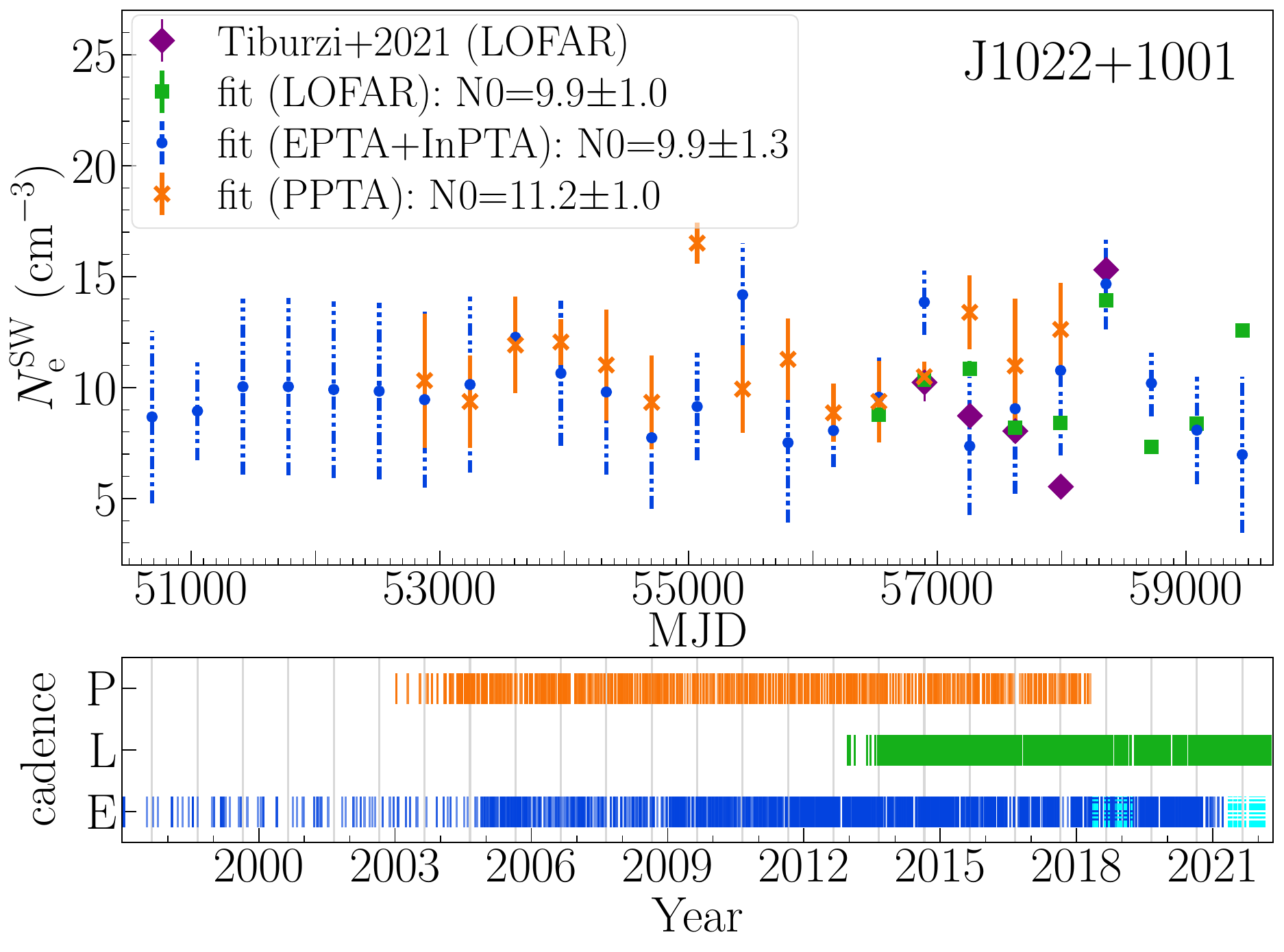} 
    \caption{The yearly SW amplitudes for PSR J1022$+$1001 are shown, resulting from the EPTA+InPTA (blue dots), PPTA (orange crosses) and LOFAR (green squares) data, as well as the corresponding values presented in T21 (purple diamonds). The cadences of the observations used are also shown, where `E' stands for the EPTA+InPTA data, with the cyan lines highlighting the InPTA data specifically; `P' stands for the PPTA data, and `L' for the LOFAR data. Note that the two x-axes are the same, but presented in both MJD and Year for convenience.}
    \label{fig: J1022_realruns}
\end{figure}

\begin{figure}
    \centering
	\includegraphics[width=\linewidth]{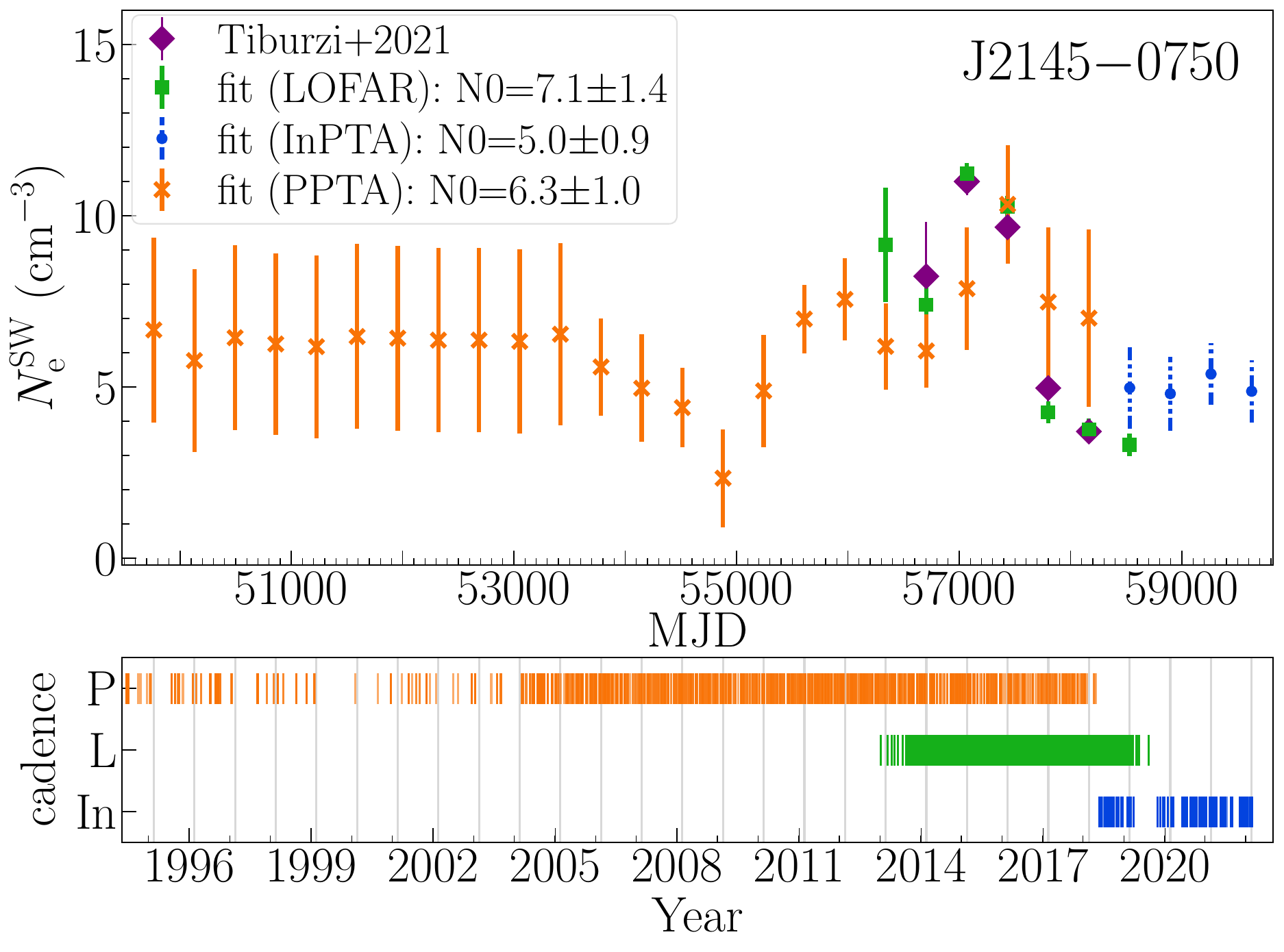} 
    \caption{The yearly SW amplitudes for PSR J2145$-$0750 are shown, resulting from the InPTA (blue dots), PPTA (orange crosses) and LOFAR (green squares) data, as well as the corresponding values presented in T21 (purple diamonds). The cadences of the observations used are also shown, where `In' stands for the InPTA data, `P' for the PPTA data, and `L' for the LOFAR data. Note that the two x-axes are the same, but presented in both MJD and Year for convenience.}
    \label{fig: J2145_realruns}
\end{figure}

Not all telescope datasets were available for each of the three pulsars chosen for this analysis; the data used is summarised in Tables~\ref{tab: realdataused} and~\ref{tab: NESWunc}. All datasets are run through the same pipeline, fitting for deterministic parameters, the SW model, white noise, achromatic red noise, and DM noise simultaneously. We compare the yearly SW amplitudes estimated from each separate dataset, for each pulsar. This is shown in Figs.~\ref{fig: J0030_realruns},~\ref{fig: J1022_realruns}, and~\ref{fig: J2145_realruns}, while the mean values, as well as the median and minimum of the SW amplitude uncertainties for each dataset are also presented in Table~\ref{tab: NESWunc}. The values of SW amplitudes from the independent LOFAR-data analysis from Fig.~6 of T21 are also included here for comparison, but are mostly discussed in the following Section~\ref{sec: compareLOFAR}.
\begin{table}
    \centering
    \caption{Table summarising the uncertainties in the measured SW amplitudes ($N_e^\mathrm{SW}$), as well as the mean values ($N_0$) from the real datasets of the three pulsars discussed, as shown in Figs.~\ref{fig: J0030_realruns},~\ref{fig: J1022_realruns}, and~\ref{fig: J2145_realruns}. The median and minimum values of these measured uncertainties (u.) are quoted in columns 3 and 4, respectively. The values in square brackets represent the median estimated only using `newer' data, where relevant, i.e. after 2004 for PPTA, and after 2005 for EPTA. Column 5 summarises the mean SW amplitude values, with the values in brackets representing one standard deviation.}
    \label{tab: NESWunc}

    \begin{tabular}{ccccc}
        \multirow{2}{*}{PSR} & \multirow{2}{*}{Dataset} & {median u.} & {min u.} & {$N_0$}\\
         & & [$\mathrm{cm}^{-3}$] & [$\mathrm{cm}^{-3}$] & [$\mathrm{cm}^{-3}$]\\
        \hline
        \hline
       \multirow{2}{*}{J0030$+$0451} & EPTA & 2.1 & 1.0 & 8.8(1.0)\\
         & LOFAR & 0.4 & 0.2 & 9.5(1.0)\\
        \hline
         & (E+In)PTA & 3.3 [2.4] & 1.5 & 9.9(1.3)\\
       J1022$+$1001 & PPTA & 2.0 [1.9] & 0.7 & 11.2(1.0)\\
         & LOFAR & 0.3 & 0.2 & 9.9(1.0)\\
        \hline
         & PPTA & 2.4 [1.5] & 1.0 & 6.3(1.0)\\
       J2145$-$0750 & InPTA & 1.0 & 0.9 & 5.0(0.9)\\
         & LOFAR & 0.3 & 0.3 & 7.1(1.4)\\
    \end{tabular}
\end{table}

Firstly, the results for PSR J0030$+$0451 are shown in Fig.~\ref{fig: J0030_realruns}.
There is agreement within uncertainties between the results of the EPTA and LOFAR datasets. As is perhaps expected, the much-lower frequency LOFAR data performs significantly better in estimating the SW amplitudes.
%The EPTA observations generally improved in the precision of the SW amplitude measurements, as the uncertainties reduced with more recent solar conjunctions. 

Secondly, Fig.~\ref{fig: J1022_realruns} shows the results for PSR J1022$+$1001. 
In general, the SW amplitude measurements are consistent between the different observing systems. While the EPTA dataset is a few years longer than the others, the early data are sparse enough that only rough estimates of the amplitudes are possible. Broadly, the PPTA data allow for more precise measurements of the SW than the EPTA data, likely owing to the additional lower-frequency and wider-bandwidth observations in the PPTA dataset. Further, the LOFAR data are seen to be much more sensitive to these measurements than the PPTA data. The collection of InPTA data has started relatively recently, and as such only two amplitude measurements are supplemented by this in the EPTA+InPTA data combination; these are the last data point and that near MJD\,58360. The increased sensitivity of the amplitude measurement which included both EPTA and InPTA data for the solar conjunction near MJD\,58360 hints at the advantage of the additional InPTA data.
%The last solar conjunction of this pulsar was in practice estimated only from InPTA data, showing a reduced precision compared to recent EPTA results despite the lower observing frequency provided by InPTA, indicative to the lack of data near the solar conjunction. Nevertheless, 

Thirdly, the SW measurements for PSR J2145$-$0750 are shown in Fig.~\ref{fig: J2145_realruns}. 
The early data (before 2004) from the PPTA represent archival data with a low cadence, such that no variation away from the mean-amplitude SW can be measured. The precision of the LOFAR measurements is significantly better than that of the PPTA measurements in the case of PSR J2145$-$0750 as well. We also note that the recent, roughly 4 years of InPTA data show an improved precision compared to the PPTA dataset for this pulsar, likely due to their high-cadence simultaneous observations.

The observed dip in SW amplitude occurring around the year 2009 (around MJD\,55000) for PSR J2145$-$0750 is coincident with the independently observed low in solar activity \citep[e.g.][]{SolarCycle}, such that our measurements appear to track the broad behaviour of the solar cycle. The same dip can also be seen, in fact, for one solar conjunction in PSR J0030$+$0451; however, this is not obviously present in the results of PSR J1022$+$1001. This suggests that, on one hand, the SW signature in pulsars further away from the ecliptic plane is likely to follow the broad 11-yr solar cycle behaviour. On the other hand, the shape of the variation away from the mean seems harder to predict for pulsars of very low ecliptic latitude (such as PSR J1022$+$1001). This may be caused, for example, 
by the slow SW contribution dominating the fast SW around the solar equator, even during the solar cycle maxima, thus showing less of the variation between the two solar cycle stages. The presence of persistent streamers near the solar equator could also induce the kind of DM variations seen in pulsars at very low heliospheric latitudes. Furthermore, the mean SW amplitude of observations of PSR J1022$+$1001, of $\sim\!(10 \pm 1)\,\mathrm{cm}^{-3}$ is significantly larger than that of PSR J2145$-$0750, of $\sim\!(6 \pm 1)\,\mathrm{cm}^{-3}$. The very similar range of observing times for these two pulsars suggests that this divergence may be due, as before, to the different ecliptic latitudes, and the corresponding heliospheric latitudes and magnetosphere areas probed. Therefore we conclude that, in general, a global fit describing the SW properties of multiple pulsars is likely to be advantageous only if the diverse regions of the SW probed by pulsars at varying heliospheric latitudes are carefully considered.

\subsubsection[LOFAR results: Comparison with Tiburzi et al. (2021)]{LOFAR results: Comparison with \citet{Tiburzi2021}} \label{sec: compareLOFAR}

Finally, the results of our pipeline are compared with the independent analysis in T21, using the same observational LOFAR data available for PSRs J0030$+$0451, J1022$+$1001, and J2145$-$0750. For our results, the data are processed through our pipeline, as with all the previous analyses presented in this work, and the yearly SW amplitudes are estimated.
The full DM series are also reconstructed from the fitted parameters, namely as the sum between the SW model, the chromatic red-noise power-law Gaussian process, and the deterministic DM model; the latter is expressed as a polynomial
\begin{equation}
    \mathrm{DM}_\mathrm{det}(t) = \mathrm{DM}_0 + \mathrm{DM}_1 \times t + \mathrm{DM}_2 \times \dfrac{t^2}{2}, 
\end{equation}
where $t$ is a ToA expressed with respect to a chosen epoch (`DMEPOCH' in \textsc{tempo2}), and the three polynomial coefficients ($\mathrm{DM}_0$, $\mathrm{DM}_1$, and $\mathrm{DM}_2$) are equivalent to the `DM', `DM1', and `DM2' fitting parameters in \textsc{tempo2}.

Conversely, the analysis in T21 was based on first obtaining a DM value per observation; for more details on how this was performed, we refer the reader to \citet{Tiburzi2019, Tiburzi2021}. To estimate the SW from the DM series, a reference value (equivalent to $\mathrm{DM}_0$) was subsequently subtracted from the DM series. The remaining contributions of the IISM and SW were disentangled and simultaneously modelled using a Bayesian framework, for each specific segment of data corresponding to a solar conjunction. The IISM contribution was modelled as a cubic polynomial for each solar-conjunction segment, and continuity between different segments was insured. In short, the total DM series in the T21 analysis was modelled as a sum between the reference value $\mathrm{DM}_0$, the IISM cubic-polynomial, and the spherical SW of yearly variable amplitude. Note that the DM series used here for comparison with our full reconstructed DM is the one initially obtained from the observations by T21, prior to the SW analysis.

\begin{figure*}
    \centering
    \includegraphics[width=0.9\textwidth]{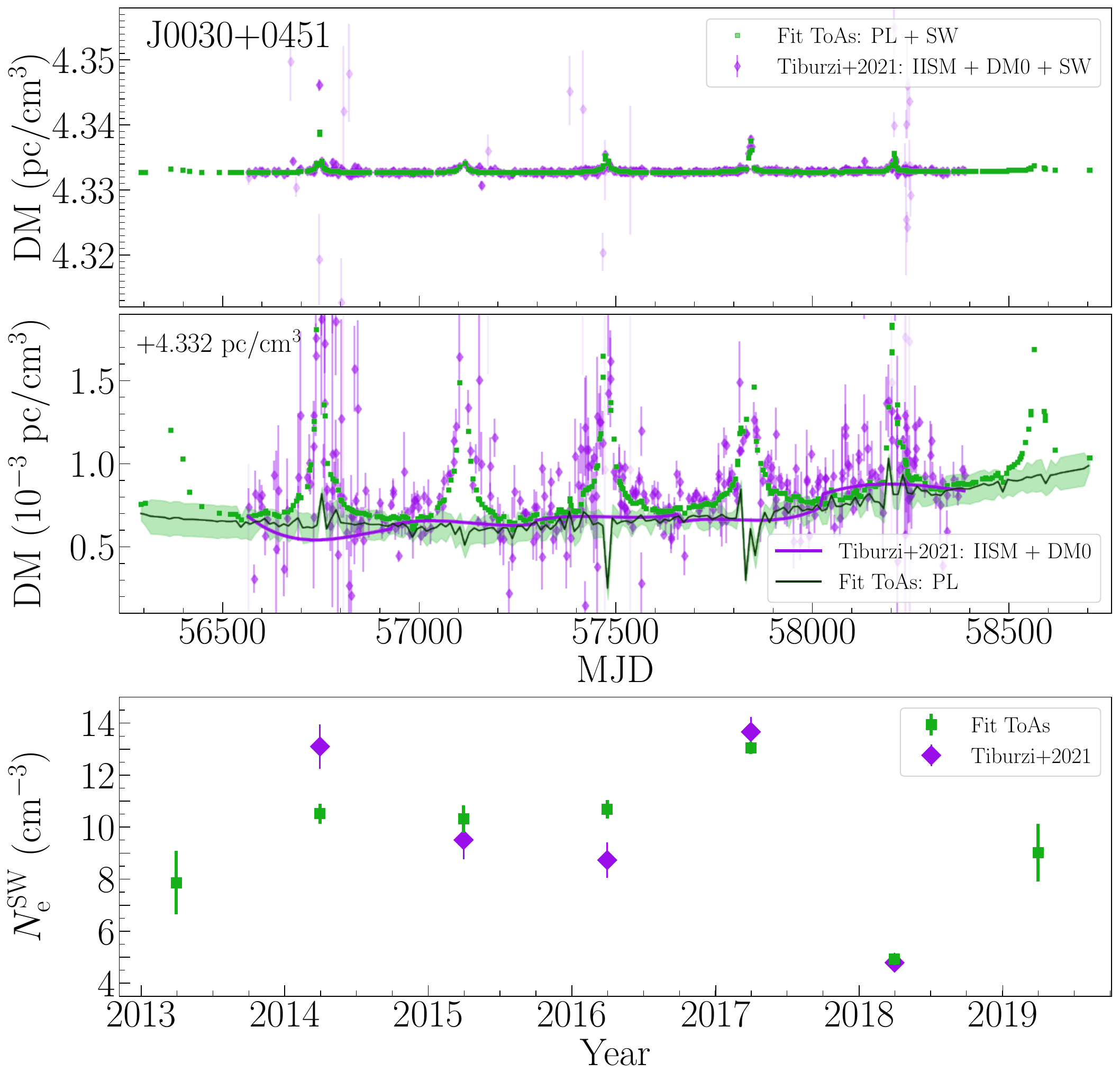}
    \caption{The recovered DM series and yearly SW amplitudes from our analysis of the LOFAR ToAs (green) are compared with the same quantities as presented by T21 for the same LOFAR data (purple), for PSR J0030$+$0451. While the top plot shows the broad picture of the DM series, the middle plot is a zoomed in version of the same data for a clear comparison. For both our results (in green), and the T21 results (in purple), the data points represent the full estimated DM series ---which includes the polynomial terms, the Gaussian-process power-law (`PL'), and the SW. The continuous lines represent only the estimated IISM contribution; this was fitted as an average $\mathrm{DM}_0$ plus multiple consecutive cubic terms in T21, and as a quadratic plus a Gaussian process power-law in our analysis. The bottom plot simply illustrates the SW amplitudes, as also shown in Fig.~\ref{fig: J0030_realruns}, but only for LOFAR data.}
    \label{fig: J0030_DMcomp}
\end{figure*}

\begin{figure*}
    \centering
    \includegraphics[width=0.9\textwidth]{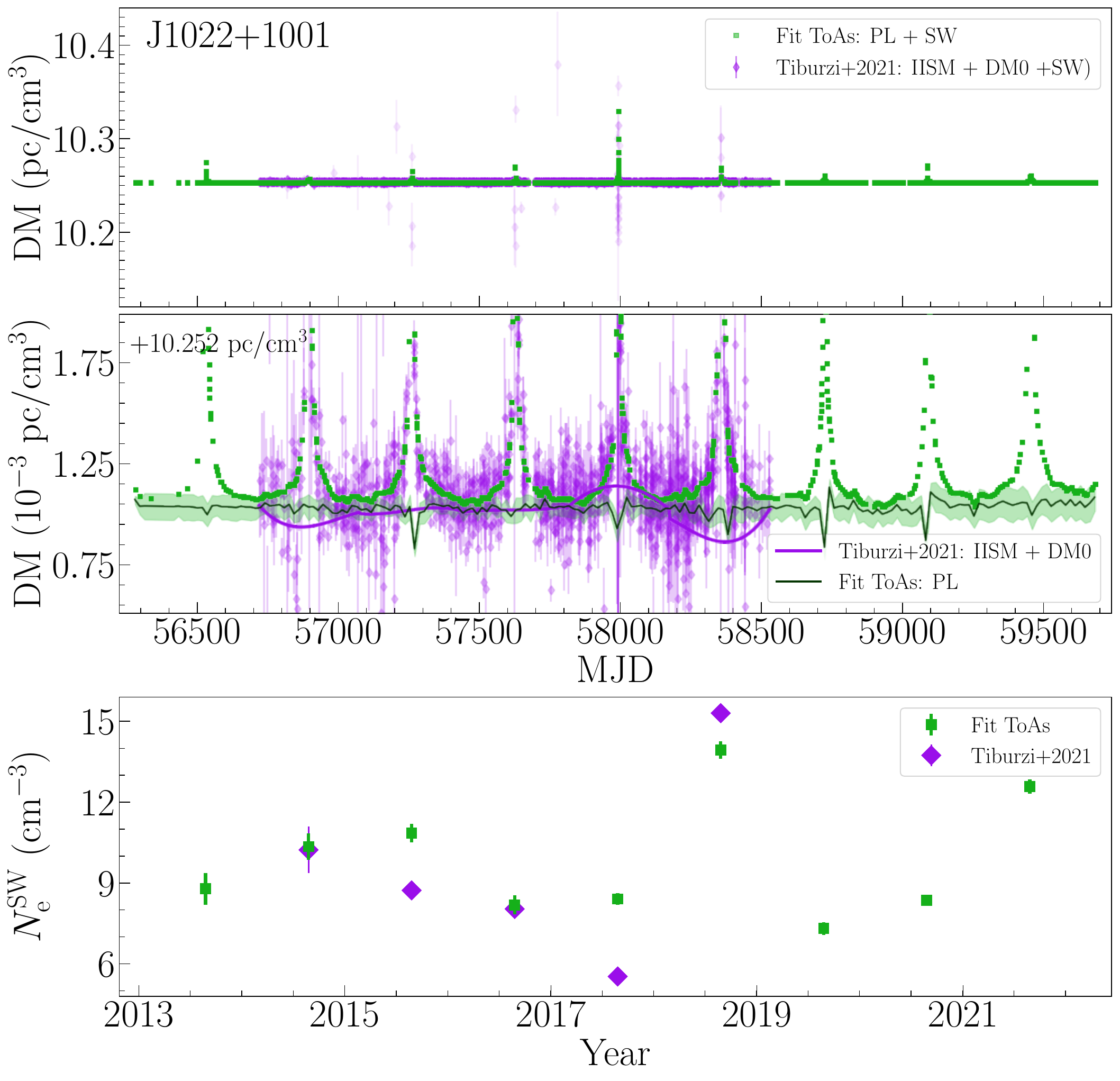}
    \caption{The same as in Fig.~\ref{fig: J0030_DMcomp}, but for PSR J1022$+$1001.}
    \label{fig: J1022_DMcomp}
\end{figure*}

\begin{figure*}
    \centering
    \includegraphics[width=0.9\textwidth]{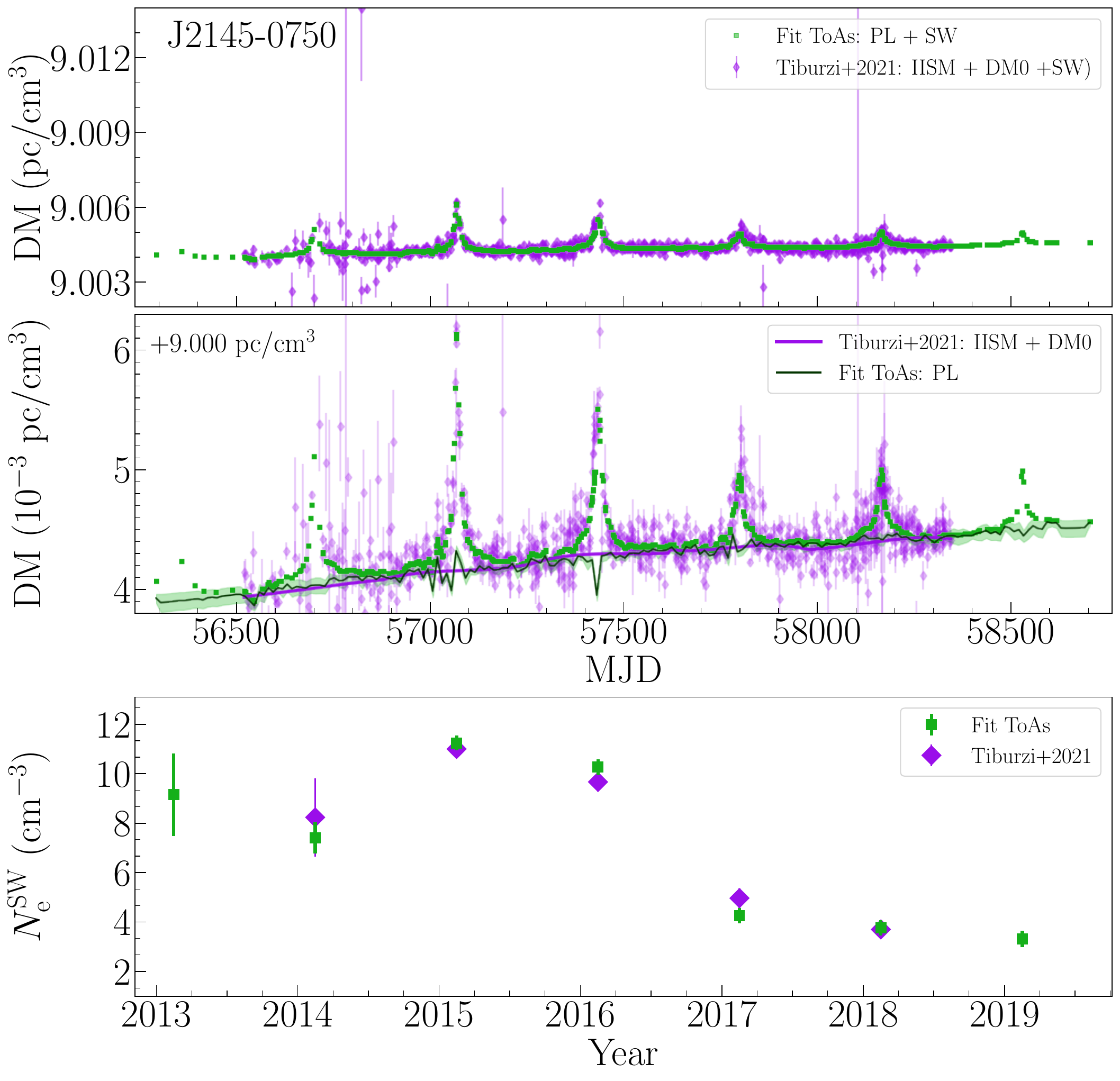}
    \caption{The same as in Fig.~\ref{fig: J0030_DMcomp}, but for PSR J2145$-$0750.}
    \label{fig: J2145_DMcomp}
\end{figure*}

Figs.~\ref{fig: J0030_DMcomp}, \ref{fig: J1022_DMcomp}, and \ref{fig: J2145_DMcomp} show the results of our analyses, as well as those of T21 for comparison, for the three PSRs J0030$+$0451, J1022$+$1001, and J2145$-$0750. Note that more recent LOFAR data were available for PSR J1022$+$1001 since the T21 work, which are included in our analysis; the additional data which appears here within one solar conjunction either side were discarded in T21 as there were not enough ToAs to provide a robust estimate by the criteria chosen in that work.
% The top plots in each figure show the total DM series in its full scale; the middle plots simply show a zoomed-in version, where the relatively small differences between the results of the two analyses are visible, both for the full DM series, and for only the `non-SW' contributions. The bottom plots show the estimated yearly amplitudes of the SW, similarly to those presented in Section~\ref{sec: compareNESW}. 

From these plots, we conclude that our estimates of the SW amplitudes are for the most part consistent with those of T21, while using an independent Bayesian approach. We note that where there are discrepancies in the SW amplitudes, they appear to be caused by the difference in the estimated IISM contribution. For example, at the solar conjunction near MJD\,58000 of PSR J1022$+$1001 (Fig.~\ref{fig: J1022_DMcomp}), our IISM estimate shows an overall flat behaviour, while the T21 analysis presented a higher, cubic-varying DM estimate. This emphasises an advantage of our pipeline with respect to the T21 method: the IISM contribution to DM is estimated for the entire dataset, rather than in segments, which in general ensures a smoother, and likely more realistic behaviour. We also note that, for the specific measurement near MJD\,58000 for PSR J1022$+$1001, the observed difference in IISM estimates of roughly $10^{-4}\,\mathrm{pc\,cm}^{-3}$ would only correspond to a direct difference in the SW amplitude ($N_\mathrm{e}^\mathrm{SW}$) of roughly $10^{-2}\,\mathrm{cm}^{-3}$; the actual measured difference is, however, of order $\sim1\,\mathrm{cm}^{-3}$. Nevertheless, we believe that, since this pulsar has a very small ecliptic latitude, the tails of the SW influence may impact the shape and peak of the SW model to a larger extent than would be trivially expected.

Lastly, it is important to acknowledge that the spherical SW model, even while allowing its amplitude to change each year, is not sufficient to fully describe the observed SW influence on these data, as also observed previously in e.g. \citet{Tiburzi2016}. This is seen in our results, as the IISM contribution to the DM clearly absorbed some asymmetric features close to the solar conjunction, which are highly likely due to the solar influence. Therefore, any analysis assuming a spherically symmetric SW, which includes that presented in this work, would not allow for an entirely robust studying of the SW influence in pulsar data. The pipeline developed and presented here, however, can create a basis that can be straight-forwardly adapted in future work to improve on this, and include various additional models, such as e.g. different piecewise components for before and after the solar conjunctions, allowing the tails of the SW variation (i.e. the pulsar ingress and egress from the Sun) to change independently of each other.

\section{Conclusions} \label{sec: conclusions}

We have implemented a computationally inexpensive, linear Gaussian-process piecewise Bayesian approach to fit pulsar ToAs for a spherical SW of yearly time-variable amplitude, simultaneously to all the other pulsar timing and noise parameters; this is available through the pulsar analysis toolkit \textsc{run\_enterprise}. We have explored the functionality of this fitting pipeline using simulations, and found that it performs as expected, and better than the currently widely used time-invariant spherical SW, particularly in data of low frequency and large fractional bandwidths.

Using simulations, we found that the current EPTA+InPTA dataset is not yet sensitive enough to measure variations of the spherical SW amplitude. However, assuming that the current observing strategy of uGMRT will continue, future EPTA+InPTA data will have increased sensitivity to the SW, such that using our time-variable fitting could improve the rms (white-noise levels) of the residuals by at least $\sim\!200\,\mathrm{ns}$ at L-band, which may help with the noise budget for extremely sensitive experiments such as the search for the GWB. Indeed, recovering the SW influence in EPTA+InPTA pulsar data will depend on the cadence of observations near the solar conjunctions, but a study of these effects is beyond the scope of this work, and is left as a future investigation.

We also applied our pipeline to real data of three pulsars, i.e. J0030$+$0451, J1022$+$1001, and J2145$-$0450, that are known to show the influence of the SW in their ToAs, and were also part of the previous study by T21. The SW amplitudes found from individual fitting of pulsar data from the EPTA+InPTA and/or the PPTA were compared with the results from LOFAR data through our pipeline, as well as the independent results from T21. This showed that SW amplitudes found from different datasets were mostly consistent with each other and that, as expected, where lower-frequency and wider-bandwidth data were present, the uncertainties were reduced. Furthermore, the variation in the fitted SW amplitudes for the pulsar of the highest ecliptic latitude in this study (J2145$-$0450; $\mathrm{elat} = 5.31^\circ$) roughly followed the 11-yr solar cycle; while for the pulsar of very low ecliptic latitude (J1022$+$1001; $\mathrm{elat} = -0.06^\circ$), the amplitude variation generally did not seem to correlate with the long-term solar cycle. This hints that a global SW fit may be more beneficial if e.g. it is performed in slices of ecliptic latitude. A larger study would clarify if this is indeed a wide systematic effect.

The DM series, including the SW effect, were also compared for the same LOFAR data as recovered from our pipeline and as used in T21. We found that where there was a difference between our SW amplitude and that in T21, this was likely due to the estimated IISM background; we believe our simultaneous fitting of deterministic and noise pulsar components produces a more plausible shape for this background than the consecutive cubic fits as used in T21. However, we conclude that even with the addition of the time-variable amplitude, the spherical SW model is not enough to fully account for the SW influence in LOFAR data. The left-over SW absorbed in the background DM series suggests that perhaps using an asymmetric model, with different components pre- and post-solar conjunction may be beneficial. The piecewise framework presented in this work could be modified for this, or for an alternate way to further study SW models.

\section*{Acknowledgements}
% Pulsar research at Jodrell Bank is supported by a consolidated grant from the UK Science and Technology Facilities Council (STFC). 
ICN was supported by the STFC doctoral training grant ST/T506291/1. The authors thank the InPTA collaboration for providing the InPTA DR1 dataset.

Part of the EPTA data used in this work is based on observations with the 100-m telescope of the Max-Planck-Institut f\"{u}r Radioastronomie (MPIfR) at Effelsberg in Germany. Pulsar research at the Jodrell Bank Centre for Astrophysics and the observations using the Lovell Telescope are supported by a Consolidated Grant (ST/T000414/1) from the UK's Science and Technology Facilities Council (STFC). The Nan{\c c}ay radio Observatory is operated by the Paris Observatory, associated with the French Centre National de la Recherche Scientifique (CNRS), and partially supported by the Region Centre in France. We acknowledge financial support from ``Programme National de Cosmologie and Galaxies'' (PNCG), and ``Programme National Hautes Energies'' (PNHE) funded by CNRS/INSU-IN2P3-INP, CEA and CNES, France. We acknowledge financial support from Agence Nationale de la Recherche (ANR-18-CE31-0015), France. The Westerbork Synthesis Radio Telescope is operated by the Netherlands Institute for Radio Astronomy (ASTRON) with support from the Netherlands Foundation for Scientific Research (NWO). The Sardinia Radio Telescope (SRT) is funded by the Department of University and Research (MIUR), the Italian Space Agency (ASI), and the Autonomous Region of Sardinia (RAS) and is operated as a National Facility by the National Institute for Astrophysics (INAF). 

The LOFAR data used in this paper is based on data obtained with: i) the German stations of the International LOFAR Telescope (ILT), constructed by ASTRON \citep{vHaarlem2013} and operated by the German LOng Wavelength (GLOW) consortium (\href{https://www.glowconsortium.de/}{glowconsortium.de}) during station-owners time and proposals LC0\_014, LC1\_048, LC2\_011, LC3\_029, LC4\_025, LT5\_001, LC9\_039, LT10\_014; LC12\_019; LC13\_007; LT14\_006 ii) the LOFAR core, during proposals LC0\_011, DDT0003, LC1\_027, LC1\_042, LC2\_010, LT3\_001, LC4\_004, LT5\_003, LC9\_041, LT10\_004, LPR12\_010; iii) the Swedish station of the ILT during observing proposals carried out from May 2015 to January 2018. We made use of data from the Effelsberg (DE601) LOFAR station funded  by the Max-Planck-Gesellschaft; the Unterweilenbach (DE602) LOFAR station funded by the Max-Planck-Institut für Astrophysik, Garching; the Tautenburg (DE603) LOFAR station funded by the State of Thuringia, supported by the European Union (EFRE) and the Federal Ministry of Education and Research (BMBF) Verbundforschung project D-LOFAR I (grant 05A08ST1); the Potsdam (DE604) LOFAR station funded by the Leibniz-Institut für Astrophysik, Potsdam; the Jülich (DE605) LOFAR station supported by the BMBF Verbundforschung project D-LOFAR I (grant 05A08LJ1); and the Norderstedt (DE609) LOFAR station funded by the BMBF Verbundforschung project D-LOFAR II (grant 05A11LJ1). The observations of the German LOFAR stations were carried out in the stand-alone GLOW mode, which is technically operated and supported by the Max-Planck-Institut für Radioastronomie, the Forschungszentrum Jülich and Bielefeld University. We acknowledge support and operation of the GLOW network, computing and storage facilities by the FZ-Jülich, the MPIfR and Bielefeld University and financial support from BMBF D-LOFAR III (grant 05A14PBA), D-LOFAR IV (grants 05A17PBA and 05A17PC1), and D-LOFAR 2.0 (grant 05A20PB1), and by the states of Nordrhein-Westfalia and Hamburg. We acknowledge support from Onsala Space Observatory for the provisioning of its facilities/observational support. The Onsala Space Observatory national research infrastructure is funded through Swedish Research Council grant No 2017-00648. 
%%%%%%%%%%%%%%%%%%%%%%%%%%%%%%%%%%%%%%%%%%%%%%%%%%
\section*{Data Availability}
The EPTA DR2 data underlying the work in this paper are available at \href{https://doi.org/10.5281/zenodo.8164424}{https://doi.org/10.5281/zenodo.8164424}, as per \citet{EPTA2023I}. The InPTA DR1 data are available at \href{https://github.com/inpta/InPTA.DR1}{https://github.com/inpta/InPTA.DR1}, as per \citet{Tarafdar2022}. The PPTA DR2 data are available at \href{https://doi.org/10.25919/cx59-a798}{https://doi.org/10.25919/cx59-a798}, as per \citet{Reardon2021}. The data from the LOFAR core can be downloaded from the Long Term Archive (LTA; \href{https://lta.lofar.eu/}{lta.lofar.eu}). The LOFAR data taken with the GLOW stations are accessible upon demand by emailing any of the first three authors.

\bibliographystyle{mnras}
\bibliography{bibliography} % if your bibtex file is called example.bib

%\begin{thebibliography}{99}
%\end{thebibliography}

%%%%%%%%%%%%%%%%%%%%%%%%%%%%%%%%%%%%%%%%%%%%%%%%%%

%%%%%%%%%%%%%%%%% APPENDICES %%%%%%%%%%%%%%%%%%%%%

%\appendix
%\section{Using the SW fitting in the run\_enterprise code}
%%%%%%%%%%%%%%%%%%%%%%%%%%%%%%%%%%%%%%%%%%%%%%%%%%

% Don't change these lines
\bsp	% typesetting comment
\label{lastpage}
\end{document}